\documentclass[11pt]{article}
\usepackage{epsfig}
\newcommand{\nc}{\newcommand}
\nc{\be}{\begin{equation}}
\nc{\ee}{\end{equation}}
\nc{\bea}{\begin{eqnarray}}
\nc{\eea}{\end{eqnarray}}
\nc{\disp}{\displaystyle}
\nc{\ade}{\mbox{$A$-$D$-$E$}}
\nc{\calN}{{\cal N}}
\nc{\calC}{{\cal C}}
\nc{\calM}{{\cal M}}
\nc{\calS}{{\cal S}}
\nc{\phit}{\hat{\varphi}}
\nc{\chit}{\hat{\chi}}
\nc{\hcalN}{\hat{\calN}}
\nc{\hcalS}{\hat{\calS}}
\nc{\hS}{\hat{S}}
\nc{\sigmad}{\sigma^\dagger}
\nc{\psid}{\psi^\dagger}

\def\sstyle{\scriptstyle}

\font\tenmsb=msbm10
\font\sevenmsb=msbm7
\font\fivemsb=msbm5
\newfam\msbfam
\textfont\msbfam=\tenmsb
\scriptfont\msbfam=\sevenmsb
\scriptscriptfont\msbfam=\fivemsb

\def\binom#1#2{{#1\choose #2}}

\def\bra#1{\langle #1|}
\def\ket#1{|#1\rangle}
\def\e{{\rm e}}
\def\d{{\rm d}}
\def\i{{\rm i}}

\begin{document}
\title{The raise and peel model of a fluctuating interface.}

\author{Jan de Gier$^1$, Bernard Nienhuis$^2$, Paul A. Pearce$^1$ and
Vladimir Rittenberg$^{1,3}$
\\[5mm]
{\small \it $^1$Department of Mathematics and Statistics, University
of Melbourne, Parkville,}\\  {\small \it Victoria 3010, Australia} \\ 
{\small \it $^2$Universiteit van Amsterdam, Valckenierstraat 65,
1018 XE Amsterdam,} \\{\small \it The Netherlands}\\ 
{\small \it$^3$Physikalisches Institut, Bonn University, 53115 Bonn, Germany.
}}

\date{\today}
\maketitle
\begin{abstract}
We propose a one-dimensional nonlocal stochastic model of adsorption and
desorption depending on one parameter, the adsorption rate. At a
special value of this parameter, the model has some interesting
features. For example, the spectrum is given by conformal field
theory, and the stationary non-equilibrium probability distribution is
given by the two-dimensional equilibrium distribution of the ice model
with domain wall type boundary conditions. This
connection is used to find exact analytic expressions for several
quantities of the stochastic model. Vice versa, some understanding of
the ice model with domain wall type boundary conditions can be obtained by
the study of the stochastic model. At the special point we study
several properties of the model, such as the height fluctuations as
well as cluster and avalanche distributions. The latter has a long tail which
shows that the model exhibits self organized criticality. We also find
in the stationary state a special phase transition without enhancement
and with a crossover exponent $\phi=2/3$. Furthermore, we study the phase
diagram of the model as a function of the adsorption rate and find two
massive phases and a scale invariant phase where conformal invariance
is broken.  
\end{abstract}

\section{Introduction}

The structure of growing interfaces is a subject of major interest and a
characterization of the various universality classes of critical behavior
is an open question \cite{BaraS95}. We present a one-dimensional
adsorption-desorption model of a fluctuating interface (see Section
\ref{se:modeldef}), which belongs to a new universality class. (Part of
the results presented in this paper were announced in
\cite{GierNPR02}). In this model, that we call the raise and peel
model (RPM), the interface follows Markovian dynamics. The adsorption
is local (it ``raises'' the interface) but in the desorption process,
part of the top layer of the interface evaporates (one ``peels'' the
interface). The relaxation rules are such that the desorption
process takes place through avalanches which have a long tail in their 
probability distribution function (PDF). The RPM therefore shows
self-organized criticality (SOC) \cite{BakTW87,Jensen98,BenHur96}. In
Section \ref{se:modeldef} we compare in detail the RPM with the Abelian sand 
pile model (ASM) and with other growth models. In this section we
also define tiles, terraces and clusters which we will use to
characterize the interface.
 
What makes our model special is that for a fine tuning of the adsorption
and desorption rates the model is solvable. Namely, the Hamiltonian,
which describes the time evolution of the system, is given 
by a sector of an XXZ quantum chain \cite{BatchGN01,PearceRGN02}. The
spectrum of the Hamiltonian can be obtained using the Bethe Ansatz
\cite{AlcarazBBBQ} and is given by a $c=0$ logarithmic conformal field
theory (LCFT) \cite{KogN02} ($c$ is the central
charge of the Virasoro algebra). This implies that the dynamic critical
exponent $z=1$. The connection between the RPM and LCFT was presented
elsewhere \cite{PearceRGN02}. LCFT appears also in other domains of
physics such as systems with quenched disorder and the quantum Hall
effect \cite{GurL99}, lattice models with $N=2$ supersymmetry
\cite{SchouF02} and possibly string theory \cite{KogP01}. Once the
proper observables are identified the existence of a LCFT behind the
stochastic process allows in principle to find the correlation
functions of the stochastic process. 

The model is special also for a second reason which will take us
into the world of combinatorics. The stationary PDF of the system with
open boundary conditions is given in terms of weighted restricted
solid-on-solid (RSOS) paths. Since there is no detailed balance in the
stochastic process we expect it to describe a state ``far away from
equilibrium''. A number of mathematical conjectures and theorems will
allow us to show that in fact the PDF can be understood as an
equilibrium PDF defined on a special two-dimensional grid. It turns
out that there exists a conjecture \cite{BatchGN01} (conjecture I)
that the properly defined normalization factor of the PDF coincides
with the number of vertically symmetric alternating sign matrices
\cite{Robbins00,Kupe00}. Alternating sign matrices are an important
research topic in combinatorics \cite{Bress99}. There is a bijection
between vertically symmetric alternating sign matrices and the ice model
\cite{Lieb67} defined on a rectangle with special boundary conditions
\cite{Kupe00}. Another bijection relates the ice model
with special boundary conditions to a fully packed loop model
(FPLM) \cite{ElkiesKLP92,BatchBNY96,Wiel00}. A second conjecture
(conjecture II) states that the weight of an RSOS path in the PDF of
the stationary state of our model is given by the number of FPLM
configurations with the same topology \cite{PearceRGN02,RazuS01}. (In
Section \ref{se:FPL} we  will review these facts). We would like
to mention that in combinatorics it takes time to prove conjectures. 
 
The conclusion of this chain of arguments is that the weighted RSOS paths
in the expression of the PDF of the stationary state can be understood as
a uniform PDF in terms of FPLM configurations for which one can use
thermodynamics. We will make use of this possibility in Section
\ref{se:clusters}. Moreover, the connections between the stationary
state of our interface model, the ice model with special boundary
conditions and alternating sign matrices allows to gain new insights
also in the last two topics where there are many open questions
\cite{Zinn00,Propp01} and few facts are known about correlation functions
\cite{BogoPZ02,Strog02}.  
 
``Nice'' combinatorial objects (the alternating sign matrices are an
example) have the magic property that various quantities have simple
product expressions. Looking at a few cases one can conjecture exact
expressions which remain to be proven. The number of alternating sign
matrices was such an example \cite{Bress99} which took some time to be
proved. In the next sections several conjectures will
be presented and we will study their physical relevance.
 
In Section \ref{se:heights} we give a conjecture (conjecture III) for
the average size of the terraces and study numerically the size
dependence of the average height and width of the interface.
 
Section \ref{se:clusters} contains a recent conjecture (conjecture IV)
about the probability to have $k$ clusters in the stationary state for
a system of size $L$. We introduce a fugacity $\zeta$ associated to $k$ and
study the thermodynamic potential of the ensemble of clusters. The
thermodynamic potential shows the correct convexity properties, which
is a test for the correctness of the conjecture. Moreover, one obtains a phase
transition from a phase with a zero density of clusters (small values
of $\zeta$) to a phase with a finite density of clusters (large values
of $\zeta$). The phase transition takes place when the fugacity is
equal to $1$. This implies that one has a special surface phase
transition \cite{Bind83} without needing an enhancement ($\zeta>1$)
for the number of clusters. To our knowledge this phenomenon was not
seen for other systems \cite{Balian73,Burk89,Queiroz95}. One also
finds a crossover exponent \cite{Bind83} $\phi=2/3$. Since the
endpoints of clusters are given by what are usually called contacts,
from the value of $\phi$ we determine the critical exponent related to
the contact-contact two-point function.
 
In Section \ref{se:ava} we study the production of avalanches in our
model. We give a conjecture (conjecture V) for the probabilities to
have an adsorption or a desorption event for different system
sizes. We also study numerically the moments of the PDF for desorption processes
with different tile numbers (avalanche sizes). We show that for
large systems the PDF has a divergent dispersion i.e. the PDF shows a
long tail. This shows that our model is in the SOC class.
 
Section \ref{se:time} deals with the RPM when one changes the value of the
ratio of the adsorption and desorption rates away from $1$. No
analytical methods are known for this new situation. We therefore
have numerically diagonalized the Hamiltonian and used finite size
scaling in order to understand the phase structure.
 
Lastly, in Section \ref{se:conc}, we present our conclusions in which
we try to convince the reader that in spite of the fact that the model
was not inspired by a particular physical phenomenon, it has many
remarkable properties bringing together different aspects of physics
and mathematics.  
\section{The model}
\label{se:modeldef}
We consider an interface of a one-dimensional lattice of size $L+1$
($L=2n$). The non-negative heights $h_i$ obey the restricted
solid-on-solid (RSOS) rules,
\be
h_{i+1} - h_i = \pm1,\qquad h_0 = h_L = 0,\qquad h_i \geq 0.
\label{eq:heights}
\ee
Alternatively, one can describe the interface using slope variables
$s_i = (h_{i+1} - h_{i-1})/2,\; (i=1,...,L-1)$. In order to
characterize the interface, we give some useful definitions. A segment
of a configuration with endpoints at the sites $a$ and $b$ is defined
by the conditions: $h_a = h_b =h$ and $h_j > h$ for $a<j<b$. If $h=0$,
the segment is called a cluster. A terrace is an interval where the
slopes are zero for all the sites. 

\begin{figure}[h]
\centerline{
\begin{picture}(160,40)
\put(0,0){\epsfxsize=160pt\epsfbox{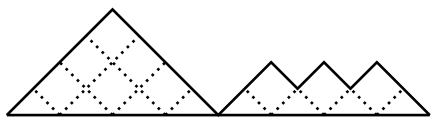}}
\end{picture}}
\caption{A configuration of the interface with two clusters.}
\label{fig:config}
\end{figure}

There are $C_n= (2n)!/((n+1)(n!)^2)$ possible configurations of the
interface. In Fig.~\ref{fig:config} we show a configuration for
$n=8,\; (L=16)$. This configuration has two clusters, a terrace of
length one separating the two clusters, one terrace of length one at
the peak of the first cluster and a terrace of length five inside the
second cluster.

The dynamics of the interface is described in a transparent way in the
language of tiles (tilted squares) which cover the area between the
interface and the substrate $(h_{2i}=0$, $h_{2i+1} =1$,
$(i=0,...,n))$. There are six tiles in the first cluster in
Fig. \ref{fig:config} and three in the second.
 
We consider the interface as separating a film of tiles
deposited on the substrate, from a rarefied gas of tiles. We are interested
to find the evolution of the interface towards the stationary state and to
study the properties  of the interface in this state.

The evolution of the system in discrete time (Monte-Carlo steps) is 
given by the following rules. With a probability $P_i= 1/(L-1)$ a tile from
the gas hits the site $i,\; (i=1,...L-1)$. Depending on the value of the slope
$s_i$ at the site $i$, the following processes can occur:
\begin{itemize}
\item[i)] $s_i=0$ and $h_i > h_{i-1}$. 

The tile hits a local peak and is reflected. 
\item[ii)] $s_i=0$ and $h_i < h_{i-1}$. 

The tile hits a local minimum. With a probability $u_{\rm a}$ the tile
is adsorbed ($h_i \mapsto h_i+2$) and with a probability $1-u_{\rm
a}$ the tile is reflected.
\item[iii)] $s_i=1$.

With probability $u_{\rm d}$ the tile is reflected after triggering
the desorption of a layer of tiles from the segment $h_{i+b} = h_i$,
i.e. $h_j \mapsto h_j-2$ for $j=i+1,...,i+b-1$. This layer contains
$b-1$ tiles. With a probability $1-u_{\rm d}$, the tile is
reflected and no desorption takes place.
\item[iv)] $s_i=-1$.

With probability $u_{\rm d}$ the tile is reflected after triggering
the desorption of a layer of tiles belonging to the segment
$h_{i-b} = h_i$, i.e. $h_j \mapsto h_j-2$ for $j=i-b+1,...,i-1$. With
a probability $1-u_{\rm d}$ the tile is reflected and no desorption
takes place.  
\end{itemize}
\begin{figure}[h]
\centerline{
\begin{picture}(160,80)
\put(0,10){\epsfxsize=160pt\epsfbox{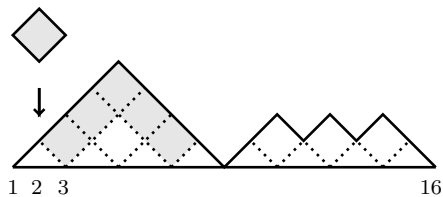}}
\put(-2,0){$\sstyle 1$}
\put(7,0){$\sstyle 2$}
\put(17,0){$\sstyle 3$}
\put(154,0){$\sstyle 16$}
\end{picture}}
\caption{A desorption event. The incoming tile at site $1$ triggers an
avalanche of $5$ tiles, which are shaded. All of the shaded tiles are
removed in the desorption event.} 
\label{fig:desorp}
\end{figure}
In our model the adsorption, which occurs on terraces, is local but the
desorption is not. To illustrate how the desorption takes place, we
show in Fig. \ref{fig:desorp} the layer of tiles which is desorbed
after a tile has hit the site $1$, there are $5$ tiles
desorbed. The same tiles are desorbed if the incoming tile would have
hit site $7$.
Desorption takes place within one cluster. We notice that
the number of tiles removed through desorption (this number is always
odd) can be of the order of the system size $L$ and therefore one can
conclude that in our model the desorption takes place through avalanches.  

Before discussing the physics of the model, we give the rates for the
continuous time evolution of the interface. At the local minima of the
interface, adsorption ($h_i \mapsto h_i+2$) takes place with a rate
$u= u_{\rm a}/u_{\rm d}$. Desorption of a segment $h_a = h_b$, i.e. $h_j
\mapsto h_j-2$ for $a<j<b$, takes place with a rate,
\be
\delta(s_a - 1) + \delta(s_b - 1). 
\label{eq:rate}
\ee
The rate $u$ is the single free parameter in the model.
 
The Hamiltonian $H$ which gives the time evolution in the vector space of
RSOS configurations, has matrix elements $H_{cd} = -r_{cd}$, where
$r_{cd}$ are the rates for the transitions $d \rightarrow c$ given
above ($\sum_c H_{cd} =0$).
The unnormalized probability $P_c(t)$ to find the system in the
configuration $c$ at time $t$ is given by the imaginary time
Schr\"odinger equation,
\be
\frac{\d}{\d t} P_c(t) = -\sum_d H_{cd}P_d(t).
\ee
Since $H$ is an intensity matrix, it has a zero eigenvalue with a trivial
bra and a nontrivial ket which gives the probabilities in the stationary
state,
\be
\begin{array}{l}
\displaystyle \bra0\,H=0,\quad
\bra0=(1,1,\ldots,1), \\[3mm]
\displaystyle
H\ket0 = 0,\quad \ket0=\sum_c P_c\ket c,\quad P_c=\lim_{t\to\infty}
P_c(t).
\end{array}
\label{eq:statdef}
\ee
In most of this paper we will consider $u=1$, Section
\ref{se:time} being an exception.  

As we are going to show in Section \ref{se:ava}, our model for $u=1$ is
of the SOC class (self-organized criticality). We here anticipate some
results and compare the present model with the Abelian sandpile model
(ASM) of Bak, Tang and Wiesenfeld \cite{BakTW87} which is a paradigm
for SOC: 
\begin{itemize}
\item[i)] In the stationary state of the RPM the balance is obtained when the
number of tiles adsorbed by the film of tiles on the substrate is equal
to the number of tiles desorbed. In the ASM the balance is obtained when
the number of grains of sand adsorbed by the film of sand is equal to
number of grains that leave the film through the boundary.
\item[ii)] The heights in the RPM can be of the order of the system size
(see Section \ref{se:heights}) therefore the RPM can also be viewed as
a growth model. The heights are finite in ASM.
\item[iii)] The RSOS paths of the RPM correspond to the recurrent
configurations of the ASM. While through toppling one configuration of
ASM is taken into one other configuration only, in the RPM one
configuration can go, with different probabilities, into several
configurations. One should add that in the ASM, topplings give a simple
physical interpretation for the avalanches. There is not yet a simple
physical mechanism for desorption transitions in the RPM.
\item[iv)] The ASM is not critical in $d=1$ \cite{Ruelle02}, but it is
in $d=2$ \cite{Ruelle01}. In both dimensions the stationary
probability distribution is uniform in the space of recurrent
states. The $d=1$ RPM is critical, the stationary probability
distribution in RPM is not uniform in the space of RSOS paths (see
Section \ref{se:FPL}). 
\item[v)] The $2+1$ ($2$ space, $1$ time) dimensional ASM has a
dynamic exponent $z=2$ \cite{IvashP98,Dhar99} and the stationary
$2$-d probability distribution describes a critical system with
correlation functions given by a $c=-2$ LCFT. At $u=1$ the $1+1$
RPM has a dynamic exponent $z=1$ and the spectrum of the
Hamiltonian is given by generic characters of a $c=0$ LCFT. The
nonuniform stationary
probability distribution in the space of RSOS paths can be obtained by
making a correspondence to the $2$-d ice model with special boundary
conditions (see Section \ref{se:FPL}). What is therefore common to the
two models is that that the stationary probability distributions which
describe phenomena ``far away from equilibrium'' are given by $2$-d
equilibrium systems. 
\item[vi)] Avalanches can be studied if one perturbs the system near the
stationary state. For the ASM the average size of the avalanche diverges
with the size of the system, whereas it stays finite for the RPM. This
is true for at least a large range of values of the parameter $u$. The
explanation is simple. We denote by $P_{\rm a}$ the probability to
have adsorption and by $P_{\rm d}$ the probability to have desorption
and take into account that one adsorbs only one tile. The average
number of tiles desorbed is,
\be
\langle T\rangle = \frac{P_{\rm a}}{P_{\rm d}},
\ee
which is finite
unless $P_{\rm d}$ vanishes as $L\rightarrow \infty$. 
At $u=1$ the probability distribution function for an avalanche of a
certain size has an algebraic fall-off implying that one has relatively
large probabilities to produce large avalanches (see Section
\ref{se:ava}). This is typical for SOC.
\end{itemize}
 
Since the RPM is also a model for interface growth it is interesting to
compare it with other models of this kind. For example in the Rouse
model of polymer dynamics \cite{Rouse53}, adsorption (desorption) takes place at the
local minima (maxima) with the same rate. One has detailed balance,
$z=2$ and the stationary state is given by a uniform PDF of RSOS
paths. This is a PDF for a directed polymer model (DPM), for which the
average height $\langle \overline{h}\rangle$ increases like 
$L^{1/2}$ \cite{Priv80}. A modification of the Rouse model by
Koduvely and Dhar \cite{KodD98} has local rates and detailed
balance but includes also adsorption (desorption) rates depending on the next
nearest neighbor heights. In this model one finds an increased value
of $z$ ($z \approx 2.5$), the average height and the width being of
the order of $L^{1/2}$. As we are going to see, in the RPM we do not have
detailed balance, locality is lost, and at least for $u=1$ one has
$z=1$ and the heights and widths increase logarithmically with the
size of the system.  
\section{RSOS, loop, and six-vertex correspondence}
\label{se:FPL}
In this section we make the connection between our model and the dense
O(1) or Temperley-Lieb loop model \cite{Pasq87,OwczaB87}, and we review
the conjectured relation of the stationary state to a model of fully
packed loops on a rectangle. This connection enables us to interpret
the non-equilibrium stationary state as an equilibrium PDF.

\begin{figure}[h]
\centerline{
\begin{picture}(370,40)
\put(0,0){\epsfxsize=160pt\epsfbox{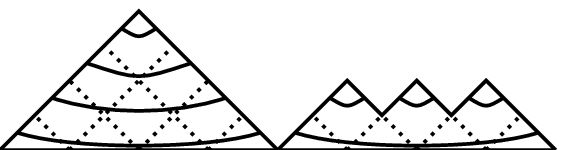}}
\put(180,0){$\leftrightarrow$}
\put(210,0){\epsfxsize=160pt\epsfbox{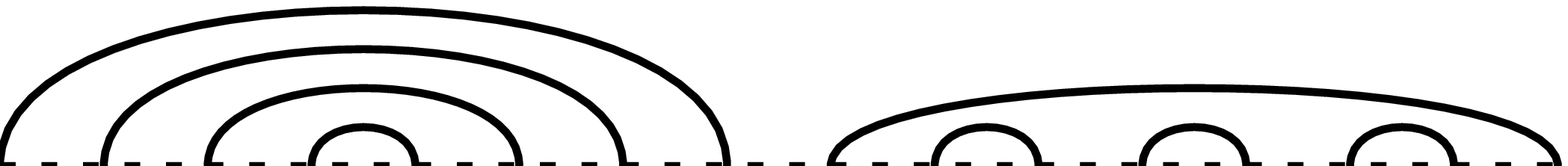}}
\end{picture}}
\caption{a) An interface with contour lines; b) Corresponding
link pattern.}
\label{fig:loops}
\end{figure}

Each RSOS path of the RPM corresponds to a ``boundary diagram'' of loops
\cite{Mart91} in the following way. On each RSOS path, draw the equal
height contour lines, as in Fig. \ref{fig:loops}a. By straightening
out the surface, keeping the contour lines and rotating the picture
around the horizontal axis, we obtain Fig. \ref{fig:loops}b. The
contour lines connect pairs of sites and Fig. \ref{fig:loops}b thus
defines a link pattern (boundary diagram). 
  
The RPM was inspired \cite{GierNPR02} by a stochastic model on the
link patterns, called the dense O(1) or Temperley-Lieb loop model
(TLLM). Namely, for $u=1$ the Hamiltonian of the RPM can be rewritten
as
\be
H = \sum_{j=1}^{L-1} (1-e_j),
\label{eq:TLham}
\ee
where the $e_j$ satisfy the Temperley-Lieb relations
\be
e_j^2 = (q+q^{-1})e_j,\qquad e_je_{j\pm1}e_j = e_j, \qquad e_je_k = e_ke_j \quad
{\rm for} \quad |j-k| > 1,
\label{eq:TLdef}
\ee
with $q=\exp(\i\pi/3)$. We remark here that the $e_j$ admit a
representation in terms of Pauli spin matrices. In that
representation, the Hamiltonian (\ref{eq:TLham}) becomes that of the
quantum XXZ spin chain with diagonal boundaries \cite{PasqS90}. This
model is integrable and quantities such as the conformal scaling
exponents can be calculated exactly, for example using Bethe Ansatz
techniques \cite{AlcarazBBBQ}. 

The $e_j$ can be pictorially represented by,
\be
e_j \quad = \quad
\begin{picture}(140,15)
\put(2,-10){\epsfxsize=135pt\epsfbox{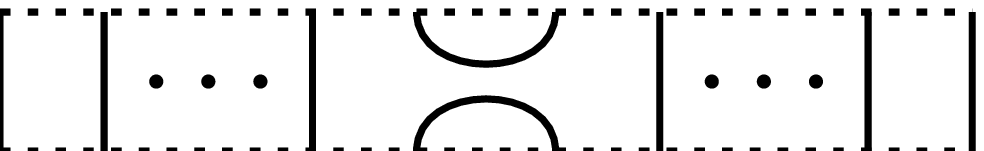}}
\put(.5,-18){$\sstyle 1$}
\put(15,-18){$\sstyle 2$}
\put(38,-18){$\sstyle j-1$}
\put(58,-18){$\sstyle j$}
\put(72,-18){$\sstyle j+1$}
\put(89,-18){$\sstyle j+2$}
\put(112,-18){$\sstyle L-1$}
\put(135.5,-18){$\sstyle L$}
\end{picture}
\hspace{10pt}\vspace{18pt}
\label{eq:monoid}
\ee
The action of $e_j$ on a link pattern of contour lines is given by
placing the graph of $e_j$ underneath that of the link pattern and
removing the closed loops and the intermediate dashed line. They allow
one to remove closed loops and contract links in composite
pictures. The action of $e_1$ on one of the link patterns for $L=6$
(corresponding to a desorption event) is for example given by, 
\be
\begin{picture}(270,40)
\put(0,0){\epsfxsize=120pt\epsfbox{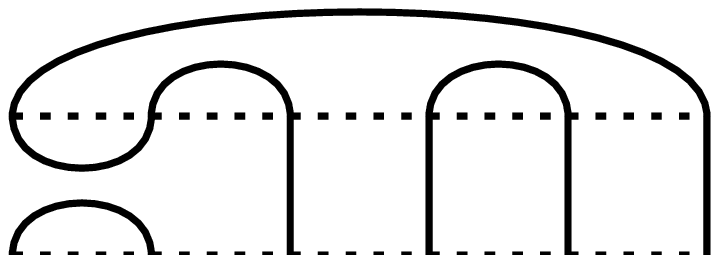}}
\put(130,10){$=$}
\put(150,10){\epsfxsize=120pt\epsfbox{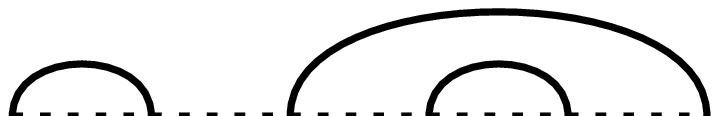}}
\end{picture}
\ee

As an example we calculate the Hamiltonian for $L=6$ on the five basis states,
\be
\begin{picture}(240,80)
\put(0,0){\epsfxsize=240pt\epsfbox{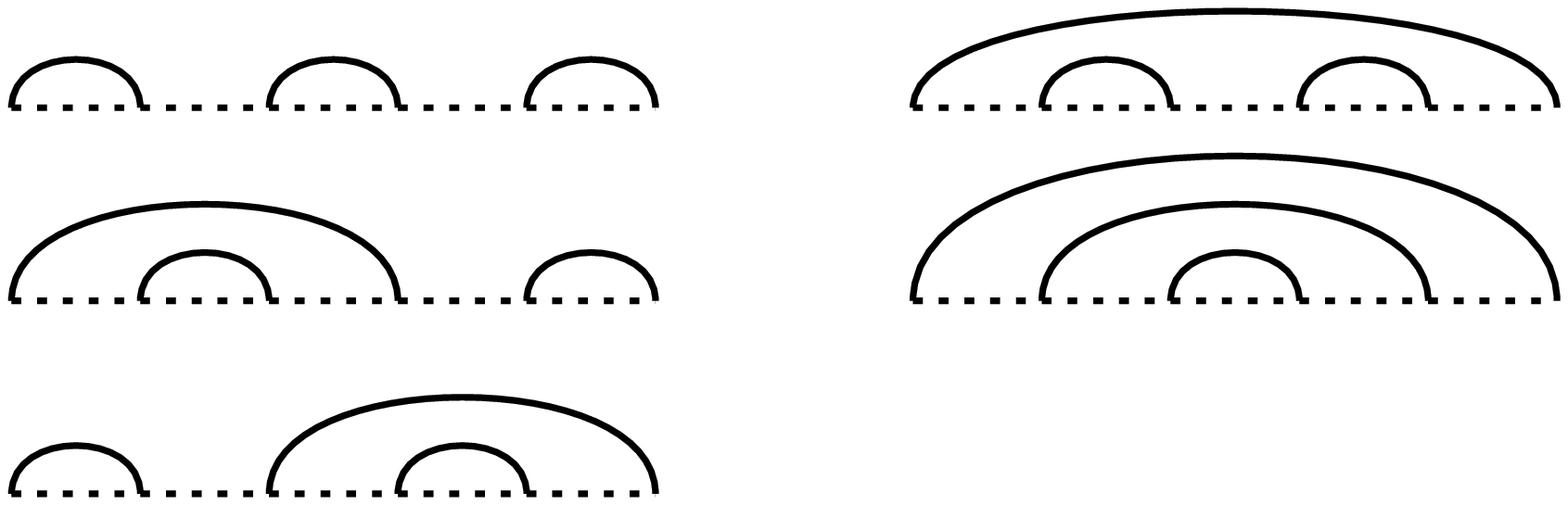}}
\put(-20,60){$1:$}
\put(-20,30){$2:$}
\put(-20,0){$3:$}
\put(120,60){$4:$}
\put(120,30){$5:$}
\end{picture}
\label{eq:basis}
\ee
and we find,
\be
H = - \left( \begin{array}{@{}ccccc@{}} 
-2 & 2 & 2 & 0 & 2 \\
1 & -3 & 0 & 1 & 0 \\
1 & 0 & -3 & 1 & 0 \\
0 & 1 & 1 & -3 & 2 \\
0 & 0 & 0 & 1 & -4 
\end{array}\right)
\ee
In the basis (\ref{eq:basis}) the stationary state $\ket0$ of $H$ is given by
\be
\ket0 = (11,5,5,4,1).
\label{eq:ket0}
\ee
Here we have chosen the smallest element to equal $1$. Using the
definitions in (\ref{eq:statdef}) we therefore find the normalization
factor to be $\bra0 0\rangle = 26$. Below we show that the
normalization acquires an extra meaning from an enumeration problem.

The RPM model for $u=1$ is thus equivalent to the Temperley-Lieb loop (TLLM)
model. For the TLLM the surprising observation was made
\cite{BatchGN01} that the normalization of the stationary state is
equal to the partition function of an equilibrium statistical
mechanics system (conjecture I). This observation has deeper
consequences which we will now briefly review.

\begin{figure}[ht]
\centerline{
\begin{picture}(100,70)
\put(0,0){\epsfxsize=100pt\epsfbox{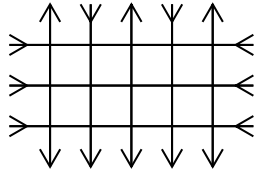}}
\end{picture}}
\caption{A $5\times 3$ rectangle with domain wall boundary conditions on the
left, right and bottom boundary and alternating arrows at the top boundary.}
\label{fig:dwbcex}
\end{figure}

For this purpose we consider a six-vertex model on a $(L-1) \times
L/2$ rectangle with boundary conditions such that the arrows on the
sides all point inward and those on the bottom boundary all point
outward (domain wall boundary conditions \cite{Kore82}). The arrows on
the top boundary alternate, see for example Fig. \ref{fig:dwbcex}. The
six-vertex configurations are in one to one correspondence with
horizontally symmetric alternating sign matrices
\cite{Robbins00,Kupe00} and can also be reformulated as configurations
of a fully packed loop model (FPLM) \cite{ElkiesKLP92,BatchBNY96,Wiel00}.

\begin{figure}[ht]
\centerline{
\begin{picture}(330,170)
\put(30,0){\epsfxsize=300pt\epsfbox{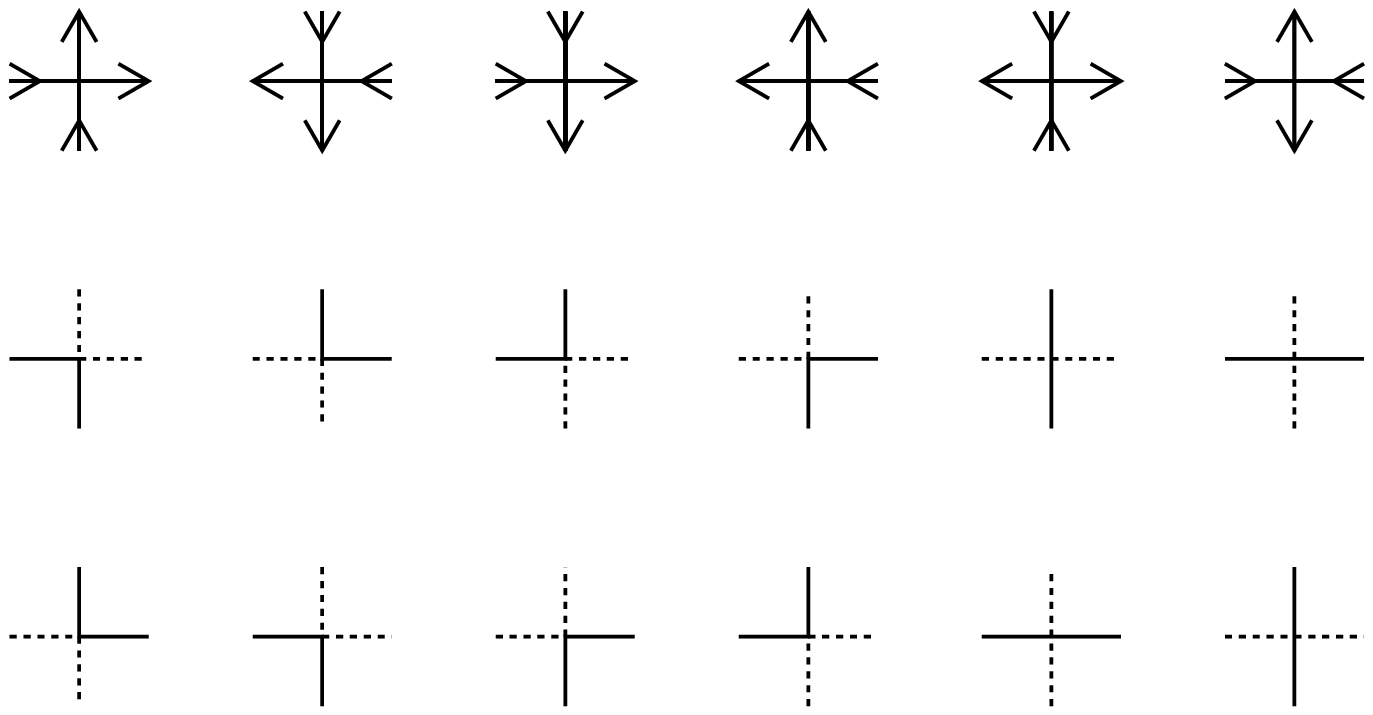}}
\put(0,78){A}
\put(0,20){B}
\end{picture}}
\caption{FPL vertices on sublattices A and B derived from the six arrow vertices.}
\label{fig:FPLlcfgs}
\end{figure}

The six-vertex configurations on the square lattice can be transformed
into fully packed loop (FPL) configurations. FPL configurations are
configurations of paths such that every site is visited by exactly
one path. We divide the square lattice into its even and odd
sublattice denoted by A and B respectively. Instead of arrows, only
those edges are drawn that on sublattice A point inward and on
sublattice B point outward, see Fig. \ref{fig:FPLlcfgs}. 

\begin{figure}[h]
\centerline{
\begin{picture}(90,50)
\put(0,0){\epsfxsize=90pt\epsfbox{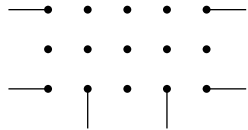}}
\end{picture}}
\caption{FPL grid corresponding to Fig. \ref{fig:dwbcex}}
\label{fig:grid}
\end{figure}

We take the vertex in the upper left corner to belong to sublattice A.
The special six-vertex boundary condition translates into a boundary
condition for the loops. Paths either form closed loops, or begin
and end on boundary sites which are prescribed by the boundary in- and
out-arrows on sublattice A and B respectively. In this way it can be
seen that the six-vertex configurations on the rectangle in
Fig. \ref{fig:dwbcex} are in one to one correspondence with FPL
configurations on the grid in Fig. \ref{fig:grid}, see also the appendix.

\begin{figure}[h]
\centerline{
\begin{picture}(340,200)
\put(0,0){\epsfxsize=340pt\epsfbox{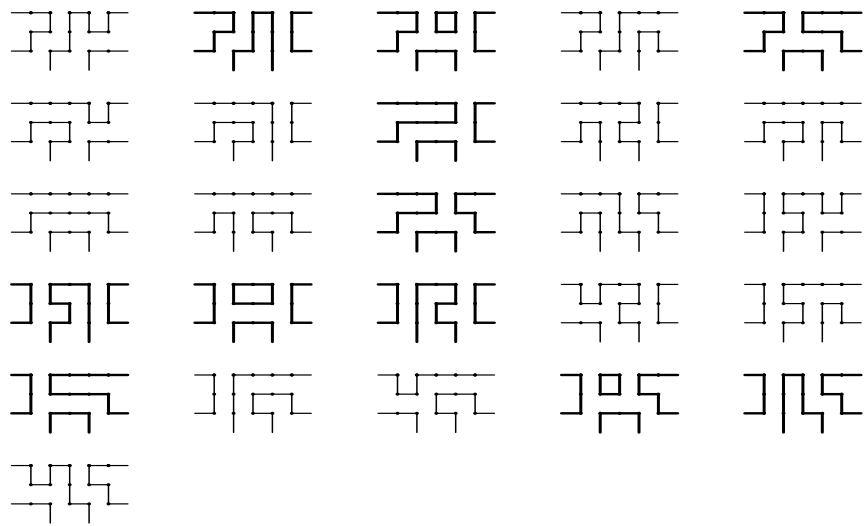}}
\end{picture}}
\caption{The $26$ FPL diagrams for $L=6$. The $11$ diagrams
corresponding to link pattern $1$ in (\ref{eq:basis}) are printed
bold.}
\label{fig:FPLpatsL6}
\end{figure}

The paths that start and end on boundary sites (we
disregard the closed loops in the bulk) define a
link pattern in the same way as did the Temperley-Lieb loops. There is
an interesting connection between the coefficients of the stationairy state
of the Hamiltonian of our model at $u=1$, or equivalently the
Hamiltonian as given by (\ref{eq:TLham}), and the enumeration of
FPL configurations on the rectangular grid: the link patterns in the
stationairy state of the TLLM and those in the FPLM appear with the
same probability \cite{RazuS01,PearceRGN02} (conjecture II).
Take for
example the stationary state for $L=6$ given by (\ref{eq:ket0}). The
FPL configurations on the $5\times 3$ rectangle are shown in
Fig. \ref{fig:FPLpatsL6}. Their total number is $26$, which is the sum
of the integers in (\ref{eq:ket0}), and they can be categorized
according to the five link patterns given in (\ref{eq:basis}). One
finds that the number of diagrams corresponding to link pattern $1$ is
$11$ (these are printed bold in Fig. \ref{fig:FPLpatsL6}), to
link pattern $2$ is $5$, to link pattern $3$ is $5$, to link pattern
$4$ is $4$ and to link pattern $5$ is $1$. 

We have just illustrated for $L=6$ the connection between the
coefficients of the stationary state of the TLLM written in the basis
of link patterns and the number of configurations of the FPLM with
that same link pattern. The connection between the stationary state
and this enumeration problem has many consequences. For example, on a
conceptual level, one understands a stationary state ``far from equilibrium''
as an equilibrium PDF. Also, the scaling properties of the RPM and the
critical properties of the FPLM are related to the scaling dimensions
of an LCFT. Furthermore, exact results obtained by studying the
properties of the enumeration problem can be used to calculate
stationary correlation functions and expectation values.

The number of six-vertex configurations on a $(L+1)\times (L+1)$ 
square with domain wall boundary conditions that are invariant under
reflection in the horizontal symmetry axis is equal to that of the
$(L-1)\times L/2$ rectangle with the special boundary condition, see
Fig. \ref{fig:dwbcex}. The total number of such configurations is
known and is equal to the number of $(L+1)\times (L+1)$ horizontally
symmetric alternating sign matrices (in existing literature more
commonly denoted as vertically symmetric alternating sign matrices)
which is given by \cite{Kupe00},  
\be
A^{\rm V}_{2n+1} = \prod_{j=0}^{n-1} (3j+2)
\frac{(2j+1)!(6j+3)!}{(4j+2)!(4j+3)!}, \qquad L=2n.
\label{eq:AV}
\ee
The leading aymptotic terms of $A^{\rm V}_{2n+1}$ are given by
\be
\ln A^{\rm V}_{L+1} = s_0 (L-1)L/2 + (s_0-\frac12\ln 2) (L-1) - \frac{5}{144}
\ln L^2 + O(1),\qquad s_0 = \ln \frac{3\sqrt{3}}{4},
\label{eq:AVasymp}
\ee
where $s_0$ is the entropy per site. The first term in
(\ref{eq:AVasymp}) is proportional to the area of the rectangle. In
the appendix we show that the second term in (\ref{eq:AVasymp}) is a
surface contribution coming from the top boundary, and not from the
other sides of the rectangle. This observation is important for the
understanding of Section \ref{se:clusters}. The surface contribution
turns out to be related to a string expectation value. 

\section{Terraces and heights in the stationary state}
\label{se:heights}
We now consider, from the point of view of growth models, various
quantities which characterize the interface in the stationary
state. In this section we look at the terraces and heights and in the
next section we will consider clusters. This separation is due to the fact
that there exists a conjecture for the cluster PDF and we will be able to
study in detail the properties of the ensemble of clusters. In Section
\ref{se:ava} we will consider the same interface from the point of
view of SOC.
 
One can make \cite{Strog} a conjecture (conjecture III) for the
fraction of the interface covered by terraces for a system of size $L$,
\be
\tau_L = \frac{1}{L-1} \sum_{j=1}^{L-1} \langle 1-|s_j|\rangle =
\frac{3L^2-2L+2}{(L-1)(4L+2)}.
\label{eq:terrace}
\ee
This conjecture was verified up to $L=18$. Eq. (\ref{eq:terrace})
implies that for large $L$, three quarters of the interface is covered
with terraces. On a terrace, half of the sites correspond to local
minima where adsorption can occur and one half to local maxima where
only reflections can occur. Desorption cannot occur on terraces. This
allows us to make an estimate of the ratio of the
probabilities to have adsorption respectively desorption: asymptotically
$P_{\rm a}/P_{\rm d} =3/2$. This relation is obtained in a different
way in Section \ref{se:ava} from another conjecture. The fact that the
results coincide is reassuring.  

The average height $\langle \overline{h}\rangle$ and interface width $w$,
which characterizes the roughness of the surface in the stationary
state, have the following definitions,
\be
\overline{h^m} = \frac{1}{L} \sum_{i=1}^{L} \lfloor h_i/2
\rfloor^m,\qquad w = \sqrt{ \langle \overline{h^2} -
\overline{h}^2 \rangle}, 
\ee
Obviously $\overline{h} =0$ for the substrate. 

\begin{figure}[ht]
\centerline{
\begin{picture}(240,190)
\put(0,20){\epsfxsize=240pt\epsfbox{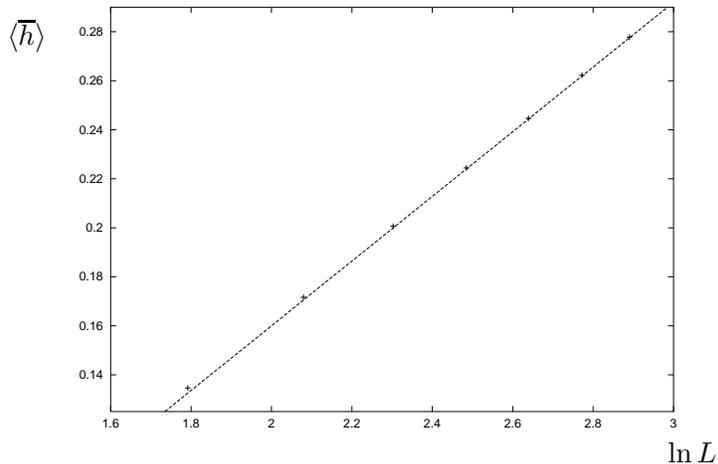}}
\put(-20,168){$\langle \overline{h}\rangle$}
\put(230,10){$\ln L$}
\end{picture}}
\caption{The $L$ dependence of the average height for
$L=6,8,\ldots,18$. The curve is given by $\langle \overline{h}\rangle
=0.132 \ln L-0.104$.} 
\label{fig:heights}
\end{figure}

We analyzed the behavior of the heights using exact data up to
$L=18$. For $L=18$ the average height has only the value $\langle
\overline{h}\rangle \approx 0.28$, which implies that the average
height increases very slowly with the size of the system. We therefore
assumed the following behavior,    
\be
\langle \overline{h}\rangle = a \ln L + b,
\label{eq:heighttrial}
\ee
and solved for $a$ and $b$ for data points corresponding to $L-2$ and
$L$. The results for successive pairs of data points are given in
Table \ref{ta:heights}. The fact that the values of $a$ and $b$ do not
change much suggests that indeed (\ref{eq:heighttrial}) may be correct. 
In Fig. \ref{fig:heights} we plot the values of $\langle
\overline{h}\rangle$ as a function of $\ln L$ together with the fit
(\ref{eq:heighttrial}) using the values for $L=18$ of Table \ref{ta:heights}.
We obtained less convincing data when fitting the heights to power laws.

\begin{table}[h]
\centerline{
\begin{tabular*}{135pt}{c|c|c}
$L$ & $a$ & $b$ \\
\hline
6 & 0.126477 & -0.092001 \\
8 & 0.128676 & -0.095941 \\
10 & 0.129859 & -0.098400 \\
12 & 0.130643 & -0.100206 \\
14 & 0.131225 & -0.101653 \\
16 & 0.131686 & -0.102870 \\
18 & 0.132067 & -0.103924 \\
\end{tabular*}}
\caption{Heights.}
\label{ta:heights}
\end{table}

\noindent
Doing a similar analysis for the widths we find that our data are
compatible with,
\be
w \sim \left(\ln L\right)^{0.35(5)}. \label{eq:widthfit} 
\ee

Because of the logs the formulae given in
(\ref{eq:heighttrial}) and (\ref{eq:widthfit}) are probably not the
last word. What is certain is that the height and the width grow
slowly with the size of the system implying that the surface is only
marginally rough. H. Hinrichsen and L. Sittler \cite{HinrS02} used
Monte Carlo simulations for our model to obtain the Family-Vicsek
\cite{Fam85} scaling function for large lattices, finding,
\be
\exp (w(L,t)^2) \sim L^{\gamma} f(t/L),\quad \gamma = 0.192 \pm 0.010.
\ee 
They confirm in this way that $z=1$ and that the width increases
logarithmically with $L$ as in (\ref{eq:widthfit}). They also
found that in the stationary state $w^2$ stays of the order of $1$
when $L$ varies between $16$ and $65536$. 

It is interesting to mention that marginally rough
surfaces (with $z=1.581$ corresponding to the directed percolation
universality class) were also encountered \cite{AlonEHM98,Hinri01} at a
critical point dividing a moving rough KPZ phase from a smooth,
massive phase. 
\section{The ensemble of clusters in the stationary state}
\label{se:clusters}
\begin{figure}[h]
\centerline{
\begin{picture}(300,80)
\put(0,10){\epsfxsize=120pt\epsfbox{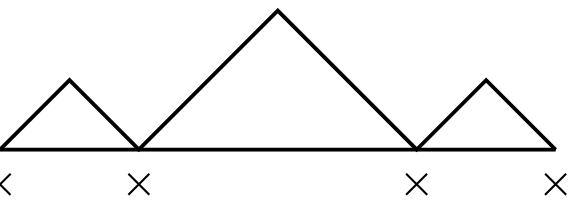}}
\put(140,30){$\sim$}
\put(180,0){\epsfxsize=120pt\epsfbox{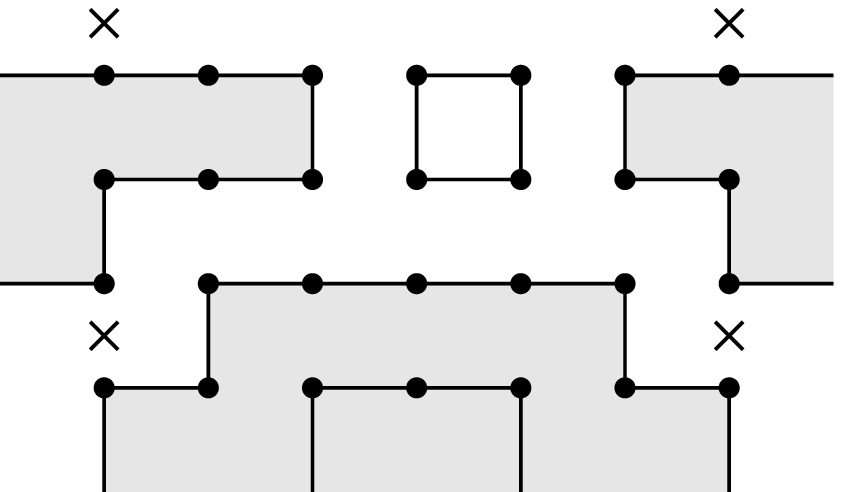}}
\end{picture}}
\caption{An interface for $L=8$. The contact points $0$, $2$, $6$ and
$8$ are denoted by $\times$. One of the corresponding FPL
diagrams, with the same contact points, is
shown on the right. The three clusters it contains are shaded.}
\label{fig:FPLclust}
\end{figure}

An obvious geometric observable to characterize a configuration of the
interface is the set of sites $j$ ($j$ even) for which $h_j=0$ 
(the sites $0$ and $L$ always belong to this set). These sites are
also called contact points. This set is
for example important to study the desorption, since two consecutive
points in this set are the endpoints of a cluster and as discussed in
Section \ref{se:modeldef} desorption takes place within one cluster
only. For this set one can define various correlation functions. Here
we will be mainly interested in the number of clusters which is
defined by
\be
k = \sum_{r=1}^n \delta(h_{2r}).
\label{eq:clusterdef}
\ee
As explained in Section \ref{se:FPL}, to
each interface, or each RSOS path, corresponds a set of FPL
diagrams. The clusters in the RSOS paths can be easily identified in the
FPL diagrams and we will take (\ref{eq:clusterdef}) also as a definition
of the clusters on FPL diagrams. The set of contact points in the interface
where $h_j=0$ maps to a corresponding set on FPL paths, see
e.g. Fig. \ref{fig:FPLclust}. 

We consider the partition function
\be
Z^{\rm FPL}_n(\zeta) = \sum_{\rm FPL\; cfgs.} \zeta^{k},
\label{eq:FPLsum}
\ee
where the summation is over FPL configurations on the $(L-1) \times L/2$
rectangular grid ($L=2n$), see Section \ref{se:FPL}, and
$\zeta=\e^\mu$ where $\mu$ is a chemical potential (or a magnetic
field). Because of the RSOS-FPL connection, the two dimensional
partition function $Z^{\rm FPL}_n(\zeta)$ can also be computed as a
in terms of weighted RSOS paths. It
is interesting to notice that the partition function $Z^{\rm 
DPM}_n(\zeta)$ for unweighted RSOS path corresponding to the stationary
state of the Rouse model (RM) \cite{Rouse53} is known \cite{Brak98}
and it was used to describe the surface critical behavior in the
directed polymer model (DPM). To see the effect of weighted RSOS
paths we will compare our results in the following to those obtained
in the DPM.

Let us define $P_n(k)$ as the probability to have $k$ clusters for a system of size
$L=2n$. The partition function (\ref{eq:FPLsum}) can then be written as
\be
Z^{\rm FPL}_n(\zeta) = {\mathcal P_n(\zeta)} A^{\rm V}_{2n+1},
\ee
where $A^{\rm V}_{2n+1}$ is given by (\ref{eq:AV}) and
\be
{\mathcal P_n(\zeta)} = \sum_{k=1}^n P_n(k) \zeta^k,\qquad {\mathcal P_n(1)} = 1.
\label{eq:Pz}
\ee
The thermodynamic potential is given for large $n$ by,
\be
\Omega^{\rm FPL}(\zeta,n) = - \ln Z^{\rm FPL}_n(\zeta) = \Omega(\zeta,n) - \ln
A^{\rm V}_{2n+1}.
\label{eq:thpot}
\ee
The leading asymptotics of the last term in (\ref{eq:thpot}) are given
in (\ref{eq:AVasymp}), and do not affect the critical properties of
the ensemble of clusters. The average number of clusters, its
second moment and the pressure can be derived from $\Omega(\zeta,n)$ via,
\bea
\langle k \rangle &=& -\zeta \frac{\partial}{\partial \zeta} \Omega(\zeta,n),
\label{eq:avkdef}\\
\langle k^2 \rangle - \langle k \rangle^2 &=& -\left(\zeta
\frac{\partial}{\partial \zeta}\right)^2 \Omega(\zeta,n), \\
p &=& -\frac{1}{2n} \Omega(\zeta,n),
\label{eq:pdef}
\eea
In order to compute the thermodynamic potential, one needs the
expression for $P_n(k)$. There is a conjecture for this expression
\cite{Gier02} (checked up to $L=18$) and its proof is an open
combinatorial problem. Let us mention that some time ago the
expression of $A^{\rm V}_{2n+1}$ was also a conjecture which was
subsequently proven. In the present paper we will accept the
conjecture for $P_n(k)$ and study its consequences for the physics of
the ensemble of clusters. This study will give an indirect
confirmation of the conjecture since convexity properties of the
thermodynamic potential are satisfied.
\subsection{A conjecture for $P_n(k)$}
In \cite{Gier02} the following conjecture (conjecture IV) for $P_n(k)$ was made,
\be
P_n(k) = k \frac{4^{n+k}}{27^n} \frac{(1/2)_{n+k}}{(1/3)_{2n}}
\frac{\Gamma(3n+2)\Gamma(2n-k)}{\Gamma(n+1)\Gamma(2n+k+2)\Gamma(n-k+1)},
\label{eq:clusterdf}
\ee
where we have used the Pochhammer symbol $(a)_n =
\Gamma(a+n)/\Gamma(a)$. The partition function (\ref{eq:Pz}) can now
be rewritten as,
\be
{\mathcal P}_n(\zeta) = A_n \zeta F_n(\zeta),
\label{eq:Pz2}
\ee
where,
\bea
A_n &=& \frac{2^{2n-1}}{27^n} \frac{(1/2)_{n}}{(1/3)_{2n}}
\frac{\Gamma(3n+2)}{n!(n+1)!(2n-1)}, \\
F_n(\zeta) &=& {}_3F_2\left( 
\begin{array}{@{}c}
2,1-n,n+3/2 \\
2-2n,3+2n
\end{array};4\zeta\right).
\label{eq:Fn}
\eea
Here we have used the definition (see e.g. \cite{Bateman}) of the hypergeometric function 
\be
{}_3F_2\left( 
\begin{array}{@{}c}
a_1,a_2,a_3 \\
b_1,b_2
\end{array};\zeta\right) = \sum_{k=1}^{\infty}
\frac{(a_1)_k(a_2)_k(a_3)_k}{(b_1)_k(b_2)_k k!}\zeta^k.
\ee
Using eqs. (\ref{eq:avkdef}) and (\ref{eq:Pz2}) we get,
\be
\langle k\rangle = 1 + \zeta \frac{F'_n(\zeta)}{F_n(\zeta)}.
\label{eq:avk}
\ee
\subsection{The thermodynamic potential of the ensemble of clusters}
To derive thermodynamic potential we have to find the asymptotics
of $F_n$. This can be achieved by starting with the
hypergeometric equation (see e.g. \cite{Bateman}) satisfied by $F_n$, 
\be
D(D+1-2n)(D+2+2n)F_n(\zeta) = 4\zeta(D+2)(D+1-n)(D+3/2+n)F_n(\zeta), 
\label{eq:hyp}
\ee
where $D=\zeta \d/\d \zeta$. We will need to distinguish the cases $\zeta<1$ and $\zeta>1$.
\begin{itemize}
\item $\zeta<1$

For small $\zeta$ we expect $F_n$ to grow as a polynomial in
$n$. Keeping only the terms with coefficients proportional to $n^2$ in
(\ref{eq:hyp}) we get
\be
F'_n(\zeta) = \frac{2}{1-\zeta} F_n(\zeta).
\label{eq:Fp}
\ee
It is clear that our assumption on $F_n$ is consistent with
(\ref{eq:Fp}) for $\zeta<1$. Using (\ref{eq:avk}) we find for the
average number of clusters,
\be
\langle k\rangle = \frac{1+\zeta}{1-\zeta}\qquad(\zeta<1).
\label{eq:avkz<1}
\ee
\item $\zeta>1$

For large $\zeta$ the main contributions will come from terms proportional
to $\zeta^n$ and thus $F_n' \sim n F_n$. From eqs. (\ref{eq:avk}) and
(\ref{eq:hyp}) we can derive the following equation for $\langle
k\rangle$, 
\be
(4\zeta-1)(D^2\langle k\rangle + 3(\langle k\rangle -1)D\langle k\rangle +
(\langle k\rangle -1)^3 = 4n^2(\zeta-1)(\langle k\rangle -1),
\ee
which for $n\rightarrow \infty$ has the following solution
\be
\langle k\rangle = \sqrt{\frac{\zeta-1}{4\zeta-1}} L\qquad (\zeta>1).
\label{eq:avkz>1}
\ee
\end{itemize}

Comparing the expressions (\ref{eq:avkz<1}) and (\ref{eq:avkz>1}) for
the number of clusters, one notices that the density of clusters
$\rho=\langle k\rangle/L$ vanishes for $\zeta<1$ (one has few large
clusters) and is finite for $\zeta>1$ (one has many small clusters). This
implies that $\zeta=1$ is the critical point of a special surface transition
\cite{Bind83}. It is interesting to note that for the DPM the critical
value of the fugacity is $\zeta_{\rm c}=2$ and not $\zeta_{\rm c}=1$ as in 
our model. The reason is simple: the critical point marks the
appearance of a finite density of clusters. This is realized more easily
for the weighted RSOS paths where the configurations with many
clusters have the largest probabilities. For a uniform distribution of
RSOS configurations, as in the DPM, one has to further increase 
the value of the fugacity to reach the critical point. This phenomenon
is the rule rather than the exception, the special transition is the
result of the enhancement obtained taking $\zeta>1$. This can be seen in
the O($n$) model ($-2\leq n<1$) \cite{Balian73,Burk89,Queiroz95} or at the collapse 
transition at the $\Theta$ point \cite{Gennes79,PrellO94,Zande98}. The
fact that in the RPM one is already at the special transition point
($\zeta_{\rm c}=1$) is probably a result of the special role played by the
boundaries in our model (see Section \ref{se:FPL}).

Let us also observe, see (\ref{eq:avkz<1}) and (\ref{eq:avkz>1}), that
for $\zeta<1$ as well as $\zeta>1$, the number of clusters 
increases with the fugacity. This is what one expects if the conjecture 
(\ref{eq:clusterdf}) is correct. From (\ref{eq:pdef}) the pressure can be
calculated for $\zeta>1$ as a function of $\rho$. We find,
\be
p = 3\sqrt{3} \left( \frac{1-\rho}{1+\rho}
\sqrt{\frac{1+2\rho}{1-2\rho}} -1\right) \approx 6\sqrt{3} \rho^3
\left( 1 + 3\rho^2 + O(\rho^3)\right).
\ee
This is an increasing function of $\rho$ which is yet another indication
of the validity of the conjecture. Notice also that for small
densities the pressure is not proportional to $\rho$ but to $\rho^3$.
%
%
%
\subsection{The cluster ensemble at criticality}
The critical behavior at a special transition is governed by a single
exponent, the cross-over exponent $\phi$ \cite{Bind83}. We expect,
\be
\langle k\rangle \sim \left\{
\renewcommand{\arraystretch}{1.4}
\begin{array}{ll}
n(\zeta-\zeta_{\rm c})^{1/\phi-1} & (\zeta>\zeta_{\rm c})\\
n^\phi & (\zeta=\zeta_{\rm c})\\
(\zeta_{\rm c}-\zeta)^{-1} & (\zeta<\zeta_{\rm c})
\end{array}\right.
\label{eq:avkscalcrit}
.
\ee
Moreover, we expect the exponent of the second moment to be twice that
of the first,
\be
\langle k^2\rangle -\langle k\rangle^2 \sim n^{2\phi}\qquad(\zeta=\zeta_{\rm c}).
\label{eq:k2}
\ee

Following Polyakov \cite{Pol70}, we expect the following scaling form
of the cluster distribution function near the critical point,
\be
P_n(k) = \frac{1}{\langle k\rangle} f(k/\langle k\rangle).
\label{eq:scalfun}
\ee
The large $x$ behavior of $f(x)$ is related to the same exponent
$\phi$ \cite{Cloiz90},
\be
\lim_{x\rightarrow\infty} f(x) \sim x^s \e^{-a x^\delta},\qquad \delta=\frac{1}{1-\phi}.
\ee
The small $x$ behavior of $f(x)$ is related to the large $n$ behavior of the probabilities $P_n(k)$,
\be
\lim_{x\rightarrow 0} f(x) = b x^\theta,\qquad 
\lim_{n\rightarrow \infty} P_n(k) = b \frac{k^\theta}{\langle
k\rangle^{1+\theta}}.
\label{eq:smallx}
\ee

For the DPM \cite{Brak98} the scaling function $f(x)$ \cite{Alex} has
the following simple expression, 
\be
f(x) = \sqrt{\frac{2}{\pi}} x \exp(-x^2/2),
\ee
from which we read off that $\phi=1/2$ and $\theta=1$. In our model the number of
clusters, calculated from (\ref{eq:avkz>1}), is equal to,
\be
\langle k\rangle = \frac{2}{\sqrt{3}} \sqrt{1-\zeta},
\ee
from which we get $\phi=2/3$ (see eq. (\ref{eq:avkscalcrit})). We have
also obtained the average number of clusters for any system size and
$\zeta=1$,
\be
\langle k\rangle = \frac{1}{3} \left(
\prod_{k=0}^{n-1}\frac{(2j+1)(3j+4)}{(j+1)(6j+1)} -1\right).
\label{eq:avkexact}
\ee
This expression was obtained by showing that it satisfies the same
recursion relation and initial conditions as its defining equation,
\be
\langle k\rangle = \sum_{k=1}^n k P_n(k),
\label{eq:kavdef}
\ee
with (\ref{eq:clusterdf}) substituted for $P_n(k)$.
Using (\ref{eq:kavdef}) one can derive (\ref{eq:avkexact}) by using an
algorithm from Zeilberger which is conveniently implemented in a
Mathematica package by Paule and Schorn \cite{PauleS95}. 
From (\ref{eq:avkexact}) we obtain the large $n$ behavior of $\langle
k\rangle$,
\be
\langle k \rangle \approx \frac{\Gamma(1/3)
\sqrt{3}}{2 \pi} (2n)^{2/3}\qquad (n\rightarrow \infty), 
\label{eq:avkasymp} 
\ee
in agreement with the second relation in (\ref{eq:avkscalcrit}). The
last of the relations of (\ref{eq:avkscalcrit}) is a consequence of
(\ref{eq:avkz<1}). The fact that all the scaling relations
(\ref{eq:avkscalcrit}) are satisfied should not be taken for granted
since our calculations are based on the conjecture
(\ref{eq:clusterdf}). 

For the fluctuations of the cluster distribution we found numerically,
\be
\langle k^2\rangle - \langle k\rangle^2 \approx 0.659(1) L^{4/3} - 0.73(1)L^{2/3}.
\label{eq:k2num}
\ee 

\begin{figure}[ht]
\centerline{
\begin{picture}(240,166)
\put(0,0){\epsfxsize=240pt\epsfbox{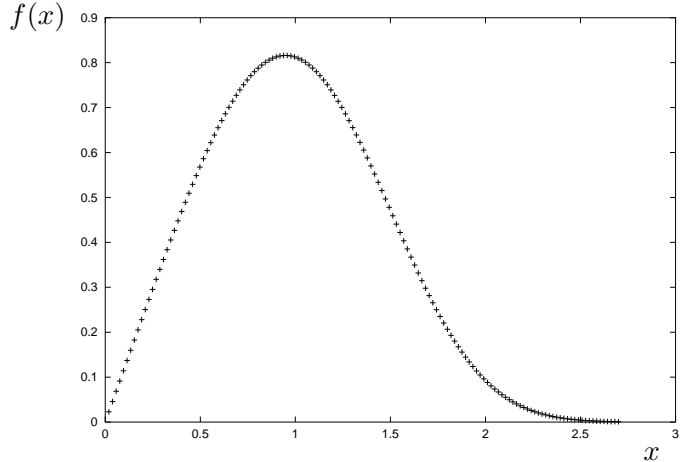}}
\put(-20,160){$f(x)$}
\put(220,-5){$x$}
\end{picture}}
\caption{The scaling function $f(x)$ obtained for $L=600$.} 
\label{fig:scalfun}
\end{figure}

The small $x$ behavior of the scaling function (\ref{eq:scalfun}) can
be determined using (\ref{eq:smallx}) and we find,
\be
f(x) \approx 3 \left(
\frac{\Gamma(1/3) \sqrt{3}}{2\pi}\right)^3 x,
\ee
which implies $\theta=1$. Finally we have determined numerically from
the conjecture (\ref{eq:clusterdf}) the scaling function
(\ref{eq:scalfun}), see Fig. \ref{fig:scalfun}. 

\subsection{$\phi=2/3$ and conformal invariance}

For isotropic systems (for DPM this is not the case), and for surface 
transitions which occur through enhancement, the crossover exponent is
related to the bulk and surface scaling dimensions \cite{Bind83,Burk89},
\be
\phi = (1-\Delta_{\rm s}) \nu = \frac{1-\Delta_{\rm
s}}{2(1-\Delta_{\rm b})},
\label{eq:phigen}
\ee
The exponent $\Delta_{\rm s}$ is related to the two-point contact correlation 
function exponent on the boundary of the halfplane. The bulk
correlation length exponent $\nu$ enters in the expression of $\phi$ because of
standard finite size scaling arguments (one uses
(\ref{eq:k2})). Eq. (\ref{eq:phigen}) can be used for the
O($n$) model ($-2\leq n<1$) and one obtains $\phi=1/2$ since $\Delta_{\rm
s}=\Delta_{\rm b}$ \cite{Burk89}, or for the collapse transition at the $\Theta$
point where one obtains $\phi=8/21$ \cite{PrellO94,Zande98} since $\Delta_{\rm
b}=1/8$ and $\Delta_{\rm s}=1/3$ \cite{Dupl87,Batch93}.

In our model, the susceptibility given by (\ref{eq:k2num}) (see also
(\ref{eq:k2})) is related to the two-point contact correlation in a
halfplane. This can be understood by looking at
Fig. \ref{fig:FPLclust}. The two sides of the rectangle can be aligned
with the bottom edge in a larger horizontal segment which for
large $L$ becomes the line which borders the halfplane. This gives the
factor $1-\Delta_{\rm s}$ in (\ref{eq:phigen}). What is more obscure
in the case of the RPM, is the finite size scaling factor $\nu$. In
our case there is no enhancement factor and no simple way to take the
system away from criticality in order to define $\nu$. If we choose to
ignore $\nu$ we have 
\be
\phi = 1-\Delta_{\rm s},
\ee
from which we derive $\Delta_{\rm s}=1/3$ which is one of the known surface
scaling dimensions \cite{ReadS01}.  
\section{Avalanches}
\label{se:ava}
In the last sections we have considered properties of the stationary
state. Here we are going to study the response of the system to small
perturbations around the stationary state. We are therefore going to
investigate the production of avalanches and show that our model exhibits
SOC. 

In order to study avalanches we consider the following processes in
discrete time (see Section \ref{se:modeldef}). In the stationary state
let a tile from the rarefied gas fall on the site $i$ with probability
$p_i=1/(L-1)$ and count how many tiles $T$ are released in the
process. This is number is zero if the tile is reflected, $-1$ if it is
adsorbed and is a positive odd number if the falling tile triggers
desorption. Repeating the process many times we can measure the probability
$R(T,L)$ to observe $T$ tiles for a system of size $L$. This PDF can
be computed from the known transition rates and the conjectured
probability distributions of RSOS configurations. 

Studying systems of different sizes $L$, we were led to the following
conjecture (conjecture V) for the probability $P_{\rm a}(L) = R(-1,L)$
to have an adsorption process and $P_{\rm d}(L) = 1-R(-1,L) - R(0,L)$
to have a desorption process,
\be
P_{\rm a}(L) = \frac{3 L(L-2)}{4(2L+1)(L-1)},\qquad P_{\rm d}(L) =
\frac{L-2}{L-1}-2P_{\rm a}(L).
\ee
This conjecture was checked up to $L=18$. The fact that one is able to obtain
simple expressions for properties of the system away from equilibrium is
remarkable since it suggests that the methods used to prove some of the
conjectures for alternating sign matrices could be extended to time
dependent properties.

Using the expressions for $P_{\rm a}(L)$ and $P_{\rm d}(L)$ we
conclude that in the large $L$ limit the average number of tiles
$\langle T\rangle$ observed in desorption increases to a maximum value
of $3/2$. This implies that, on the average, there are few tiles
desorbed. We remind the reader that in Section \ref{se:heights} using
a completely different argument we obtained the same asymptotic value
for $\langle T\rangle$. In order to analyse the properties of the
avalanches, it is convenient to write $T=2v-1$, $(v=1,2,\ldots)$ and
to consider $v$ as the size of the avalanche. Given the occurrence of
an avalanche, its size $v$ is distributed according to the PDF,
\be
S(v,L) = \frac{R(2v-1,L)}{P_{\rm d}(L)}.
\ee
If the PDF of tiles presents a long tail, i.e. if one has SOC, finite size
scaling theory (FSS) \cite{Jensen98} suggests the following form for
this PDF,
\be
S(v,L) = v^{-\tau} F(v/L^D). \label{eq:AvPDF}
\ee
One way to get the exponents of the FSS function is to consider the
moments \cite{Tebaldi99},
\be
\langle v^m\rangle = \sum_{v=1} v^m S(v,L) \sim A(m)L^{\sigma(m)},
\ee
for which one expects, 
\be
\sigma(m)=
\left\{ \begin{array}{cc}
0, & m<\tau-1 \\ 
D(m+1-\tau), & m > \tau-1
\end{array}\right. . 
\label{eq:sigma}
\ee
Since $\langle T\rangle = 3/2$, one has $\sigma(1)=0$ and $A(1)=
5/4$. In order to compute the moments, we have used the exact values
of $S(v,L)$ for $L$ up to $18$ and VBS approximants \cite{VBS}. For
$m=1.5$ one obtains $\sigma(1.5)=0$ with $A(1.5)\approx1.745 $. For
$m=2$, $\sigma(2)$ is hard to determine suggesting that the dispersion
(second moment) diverges logarithmically. In Table \ref{ta:avaexp} we
give the estimates and extrapolations for the exponents $\sigma_L(3)$,
$\sigma_L(4)$ and $\sigma_L(5)$, where 

\be
\sigma_L(m) = L \frac{\langle v^m\rangle_L - \langle v^m\rangle_{L-1}}
{\langle v^m\rangle_{L-1}}.
\ee

\begin{table}[h]
\centerline{
\begin{tabular*}{180pt}{c|c|c|c}
$L \backslash m$ & 3 & 4 & 5\\
\hline
6 & 1.31250 & 2.81250 & 5.81250 \\
8 & 1.14050 & 2.42898 & 4.59306 \\
10 & 1.05707 & 2.25429 & 4.06439 \\
12 & 1.00985 & 2.15976 & 3.78047 \\
14 & 0.98066 & 2.10312 & 3.60721 \\
16 & 0.96159 & 2.06682 & 3.49219 \\
18 & 0.94867 & 2.04244 & 3.41113 \\
\hline
$\infty$ & 0.91830 & 1.98162 & 3.02038
\end{tabular*}}
\caption{Avalanche exponents $\sigma(m)$.}
\label{ta:avaexp}
\end{table}

\begin{figure}[ht]
\centerline{
\begin{picture}(240,166)
\put(0,0){\epsfxsize=240pt\epsfbox{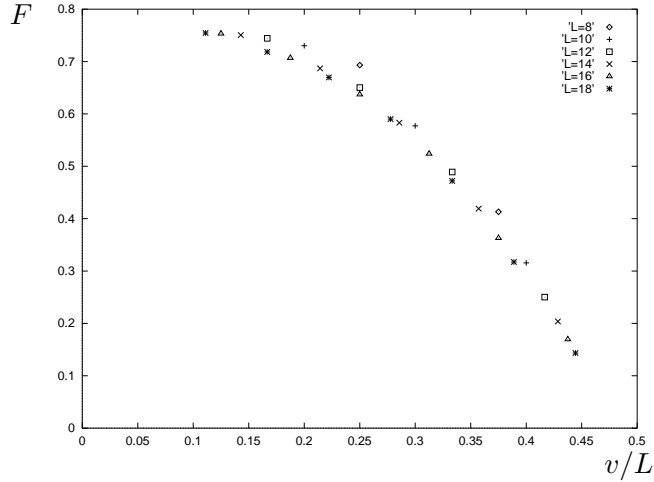}}
\put(-5,160){$F$}
\put(220,-10){$v/L$}
\end{picture}}
\caption{Avalanche scaling function $F(v/L)$. The data are obtained
for $v>1$ and $L=8,\ldots,18$.} 
\label{fig:avcoll}
\end{figure}

These results suggest that $\sigma(m)=m-2$ for $m\geq 2$ (with 
logarithmic corrections close to $m=2$). If this is indeed the case
we conclude that $D=1$ and $\tau=3$. Because of crossover effects one
cannot preclude small changes in the values of these two
exponents. The value $D=1$ was to be expected since $L$ is the single
characteristic length in our system. A consistency check was done
assuming $D=1$ in (\ref{eq:AvPDF}) to see for which value of
$\tau$ one finds a data collapse for the scaling function $F(v/L)$. Since
we have data up to $L=18$ only we cannot expect a precise value
neither of $\tau$ nor of $F$. Nevertheless as shown in
Fig. \ref{fig:avcoll}, a data collapse is visible for $\tau =3.2$, in
agreement with the value $\tau=3$ mentioned earlier. We conclude that
the PDF of tiles shows long tail and that therefore our model is in
the SOC class.
\section{The interface model away from the Temperley-Lieb rates}
\label{se:time}
In this section we study the interface model described in
Section \ref{se:modeldef} for adsorption rates $u$ different from $1$. This
study is interesting not only because we expect the physics to be
different but also from a theoretical point of view. Changing the
value of $u$, one perturbs the LCFT in a nonlocal way and it is not
clear what is the nature of the phases one obtains. In principle one
can obtain phases that are scale invariant but not conformally
invariant. Such a scenario was not yet seen, as far as we know, in
other physical systems.

Let us forget for a moment that we know what happens at $u=1$. We will
try to reason what might occur for general $u$ using physical considerations. One
has to keep in mind that the RSOS paths are confined to a triangle and
this geometry might have consequences for the phase structure.
At $u=0$ the system is not critical. The stationary state (energy
zero) corresponds to the substrate and the first degenerate excited energy
level corresponds to having a tile on top of the substrate. This
energy (energy gap) is equal to $2$ (corresponding to the rate given by 
equation (\ref{eq:rate})), and is $L-1$ times degenerate since the
tile can be desorbed from $L-1$ sites. No avalanches are produced in
the substrate.

If one slightly increases $u$ the energy gap also changes slightly (in fact it
decreases), the degeneracy of the first level is lifted and the first
Brillouin zone appears. This is the massive phase I of the model. In
this phase one expects the stationary interface to be composed mainly
of terraces, the average height $\langle \overline{h}\rangle$ and the
density of clusters should stay finite in the thermodynamic limit. The
interface is smooth in this phase and the avalanche distribution should
not show a long tail.

Increasing the value of $u$ over a critical value $u_{{\rm c},1}$ the density of
clusters should vanish and $\langle \overline{h}\rangle$ should increase
algebraically in $L$, the interface being rough. We expect this phase
to be massless and the probability distribution of the avalanches to
show a long tail,
therefore we will call this phase the SOC phase. Invoking what we know
about the physics at $u=1$ where the interface is marginally rough,
one might argue $u_{{\rm c},1}=1$.

If $u$ becomes larger than a critical value $u_{{\rm c},2}$ one expects a second
massive phase that we denote by II. Let us explain why. It is useful to
consider instead of $u$ its inverse $w=1/u$ as a variable and to take $w<1$. We
redefine the time scale such that one has a rate $1$ for adsorption and $w$ for
desorption. At $w=0$, the only RSOS path entering the stationary state is the one
corresponding to the edges of the triangle of height $L/2$. In this
configuration there is one cluster which contains the maximum number of tiles
and $\langle \overline{h}\rangle=(L-2)/8$ . Like for $u=0$, for $w=0$ we
do not have avalanches, this time because we do not have desorption. The
first excited state which corresponds to the configuration with one
tile less than in the stationary state, has an energy equal to $1$
(corresponding to the adsorption rate) and is degenerate. It turns out
that the spectrum of the Hamiltonian (including degeneracies !) at
$w=0$ is identical up to a factor $1/2$ to the spectrum of the
Hamiltonian at $u=0$. This observation deserves a more detailed
explanation which we are not going to give here. For small values of
$w$, the $L-1$ degeneracy of the first energy level is lifted giving
rise to the first Brillouin zone. Increasing the value of $w$ we span
the massive phase II up to a critical  
value $w_{{\rm c},2}=1/u_{{\rm c},2}$. In this phase, $\langle
\overline{h}\rangle$ is of the order $L$ and the density of clusters
is zero. 

\begin{figure}[h]
\centerline{
\begin{picture}(300,50)
\put(0,15){\epsfxsize=300pt\epsfbox{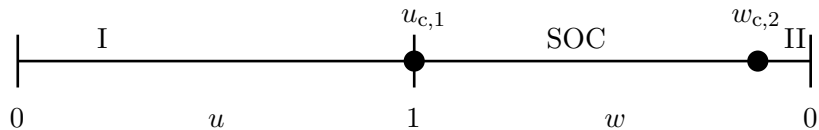}}
\put(-3,0){$0$}
\put(297,0){$0$}
\put(147,0){$1$}
\put(72,0){$u$}
\put(222,0){$w$}
\put(270,40){$w_{{\rm c},2}$}
\put(145,40){$u_{{\rm c},1}$}
\put(30,30){I}
\put(200,30){SOC}
\put(290,30){II}
\end{picture}}
\caption{Expected phase diagram} 
\label{fig:phase1}
\end{figure}

To sum up, according to this scenario, illustrated in Fig. \ref{fig:phase1}, one
expects two massive phases, one massless phase and two critical points
$u_{{\rm c},1}$ and $u_{{\rm c},2}$. 

To check this scenario we have limited ourselves to finite size scaling
studies (FSS) to find out when a system is gapless and to
determine the dynamic scaling exponent $z$. Unfortunately
analytical methods can be used for $u=1$ only.
We have diagonalized numerically the Hamiltonian up to $L=16$ and have
determined $E_1(u,L)$ and $E_2(u,L)$, the energies of the first and second
excited states as functions of $u$ and $L$ (or of $w$ and $L$).
From our
experience, when dealing with nonlocal Hamiltonians, FSS methods have some
times convergence problems and can be efficient in some cases and not in
others.

\begin{figure}[h]
\centerline{
\begin{picture}(260,180)
\put(0,20){\epsfxsize=240pt\epsfbox{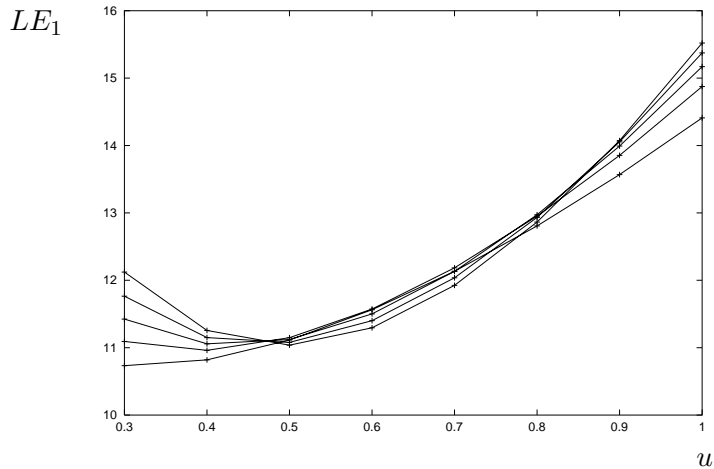}}
\put(-30,175){$L E_1$}
\put(230,10){$u$}
\end{picture}}
\caption{Scaled $E_1$ as a function of the adsorption rate $u$ for
$L=8,10,12,14$ and $16$ (the smaller systems give smaller values for
$u=1$). The curves are a guide to the eye only.}  
\label{fig:ugap}
\end{figure}

\begin{figure}[ht]
\centerline{
\begin{picture}(260,180)
\put(0,20){\epsfxsize=240pt\epsfbox{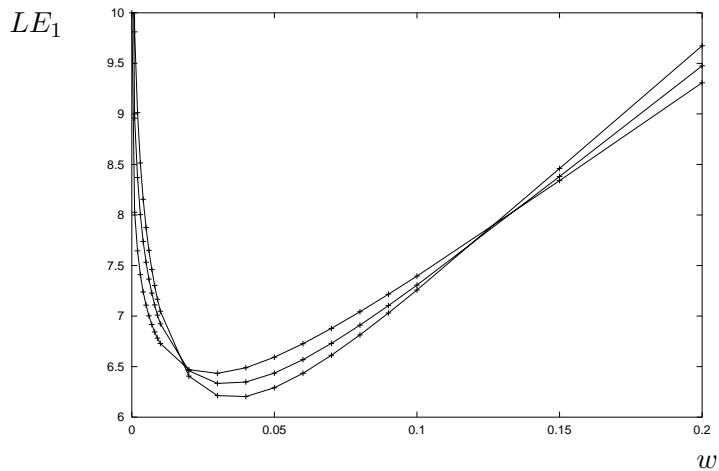}}
\put(-30,175){$L E_1$}
\put(230,10){$w$}
\end{picture}}
\caption{Scaled $E_1$ as a function of the adsorption rate $w$ for
$L=10,12$ and $14$ (the smaller systems give smaller values for
$w=0.2$). The curves are a guide to the eye only.} 
\label{fig:wgap}
\end{figure}

At
$u=1$ it is known \cite{GierNPR02} that,
\be
L E_1(1,L) = 2\pi v + o(1),\qquad  L E_2(1,L) = 3\pi v +o(1), 
\label{eq:CFTlev}
\ee
where the sound velocity $v=3\sqrt{3}/2$.
We will first assume that, like for $u=1$, in the whole SOC phase the
dynamic exponent $z=1$. In Figs. \ref{fig:ugap} and \ref{fig:wgap}
we have plotted the values of $LE_1(u,L)$ respectively $LE_1(w,L)$ for
various number of sites. If one would have a single critical point at $u=w=1$ (this
is not our scenario) one would expect crossings for values of $u$ and
$w$ close to $1$. The values of $u$ and $w$ for which one obtains
crossings for various lattice sizes should converge to $1$ for large
$L$. This is not at all what one sees. We notice that there are two 
crossings at $0.5$ and $0.85$ in the $u$ domain. The second crossing
might well converge to $1$. There are also two crossings in the $w$
domain, at $0.02$ and $0.13$. The data suggest, as expected two
massive phases, one for $u<0.5$ and another one for $w<0.02$. What
happens in between is less clear. Taking into account that we only
have results for small systems, all we could do were some consistency
checks. 

\begin{figure}[ht]
\centerline{
\begin{picture}(300,50)
\put(0,15){\epsfxsize=300pt\epsfbox{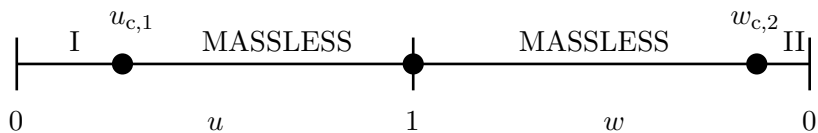}}
\put(-3,0){$0$}
\put(297,0){$0$}
\put(147,0){$1$}
\put(72,0){$u$}
\put(222,0){$w$}
\put(35,40){$u_{{\rm c},1}$}
\put(270,40){$w_{{\rm c},2}$}
\put(20,30){I}
\put(70,30){MASSLESS}
\put(190,30){MASSLESS}
\put(290,30){II}
\end{picture}}
\caption{Proposed phase diagram} 
\label{fig:phase2}
\end{figure}

One possibility is that in the whole domain $u>0.5$, $w>0.02$ one has
a massless phase with $z=1$. In order to check this possibility we have checked using
different extrapolation methods if for a given value of $u$ (respectively $w$)
the quantities $LE_1(u,L)$ converge for large $L$. For $u>0.7$ and $w$
close to $1$ the data are compatible with a massless phase with $z=1$,
although the convergence is less clean than for $u=1$ (the extrapolated
values are far away from the value for $L=16$). However, from the numerics
we cannot exclude the posibility that we only have $z=1$ at $u=1$ and a
different value, albeit very close to $1$, elsewhere. In the remaining
domain it is hard to make any clear statement because of convergence
problems. This picture invalidates already part of our scenario since
$u=1$ is in the middle of a massless phase and thus cannot correspond to
$u_{{\rm c},1}$. If that is the case, the phase diagram of
Fig. \ref{fig:phase1} should be replaced by Fig. \ref{fig:phase2}.

Can we conclude that in the domain where we have $z\approx 1$ we also
have LCFT? This possibility can be easily checked looking at the ratio
$E_2(u,L)/E_1(u,L)$ for large values of $L$. The reason we are looking
at this ratio is the following one. Because of the universality of the
amplitudes in LCFT, see (\ref{eq:CFTlev}) where now the sound velocity $v$
can be a function of $u$, this ratio should have the value $3/2$
independent of $u$ in the whole domain of the SOC
phase. In a massless phase which is not described by a LCFT this ratio can have any
value (no universality). On the other hand, in the massive phases this
ratio should be equal to $1$ since the second energy level belongs to
the same Brillouin zone as the first energy level. 

\begin{table}[h]
\centerline{
\begin{tabular*}{345pt}{c|c|c|c|c|c|c}
$L\backslash u$ & 0.01 & 0.3 & 0.9 & 1.0 & 1.111 & 1000\\ 
\hline 
8 & 1.068015 & 1.422562 & 1.419503 & 1.390377 & 1.355468 & \\
10 & 1.057506 & 1.442180 & 1.462111 & 1.425500 & 1.381065 & 
1.11059 \\
12 & 1.047632 & 1.439921 & 1.488372 & 1.446044 & 1.393992 & 
1.11237 \\
14 & 1.039483 & 1.427343 & 1.506046 & 1.459091 & 1.400681 & 
1.11164\\
16 & 1.032983 & 1.409821 & 1.518746 & 1.467896 & & \\ 
\hline
$\infty$ & 1.00691 & $-$ & 1.629253 & 1.500120 & 1.415755 &  $-$
\end{tabular*}}
\caption{Ratio $E_2/E_1$.}
\label{ta:ratios}
\end{table}

In Table \ref{ta:ratios} we give the values of the ratios for different values of $u$
and $L$ as well as (when possible) the extrapolated values for large $L$. One
notices that, as expected, for $u=1$ one gets the value $3/2$. For
$u=0.01$ (in the first massive phase) the extrapolated value is, as
expected $1$. For $u=0.3$ (also expected to be in the massive phase
I, the ratio first increases with $L$ and then decreases, probably
towards the value $1$. In the second massive phase ($u=1000$), the
values of the ratios first increase (from $L=10$ to $L=12$) and then
decrease (from $L=12$ to $L=14$). This makes any extrapolation 
procedure meaningless and for this reason we didn't give a value for
$L=\infty$ in Table \ref{ta:ratios}. Nevertheless, one can see
that for $u=1000$, the three values of the ratio are close to $1$ (the expected value). These
observations provide a supplementary check for the existence of the two
massive phases.

We now come to the most interesting result of our analysis. For $u=0.9$ and
$1.111$, assumed to be in the SOC phase, the ratios do not converge to
the value $1.5$. This implies that although one is scale invariant ($z$
close to $1$) one is not conformally invariant.  

To conclude, although based only on limited FSS studies, our results
suggest that changing the value of the rates from their Temperley-Lieb
values gives a phase diagram with two massive phases and a massless
phase with a nontrivial structure. There is one point ($u=1$)
described by a LCFT and around this point one has a domain, of a size
which is not well known, where one stays scale invariant but one doesn't have
LCFT. Hopefully Monte-Carlo simulations will clarify this picture.
\section{Conlusions}
\label{se:conc}
The Hamiltonian of the Temperley-Lieb loop model (TLLM) defined in the
space of link patterns for $q=\exp(\i\pi/3)$ (see eq. (\ref{eq:TLdef}))
is an intensity matrix giving the time evolution of a stochastic
process. As shown in \cite{ReadS01} its spectrum is given by the characters
of a $c=0$ LCFT. The same Hamiltonian corresponds to a sector of the
XXZ quantum chain with quantum group boundary conditions \cite{PasqS90}. The
first motivation in our study was to see the specific
features of a stochastic model in which one has LCFT. Mapping the link
patterns into RSOS configurations (see Section \ref{se:FPL}) we get a
model of a fluctuating interface: the raise and peel model (RPM) described in
Section \ref{se:modeldef} for $u=1$. The main observation is that this
model is simple and that the study of its physics is very
interesting. This was a pleasant surprise. The physics is not so
transparent for other stochastic models related to extensions of the
TLLM \cite{PearceRGN02,BatchGN02}.  

A particular feature of the RPM is that the PDF which gives the stationary
distribution of the RSOS configurations can be understood in terms of an
enumeration of fully packed loop (FPL) configurations on a rectangle for which
one side plays a different role than the other three. This is one of the
conjectures \cite{PearceRGN02,RazuS01} on which this paper is
based. Accepting this conjecture which was checked for small system
sizes, one concludes that the ``far away from equilibrium'' stationary
state of the one-dimensional RPM is in fact a two-dimensional
equilibrium distribution.

This is not the end of the story. The FPL configurations can be mapped
into the ice model on a rectangle with domain wall boundary conditions on
three edges and alternating arrows on the fourth (see Section
\ref{se:FPL}). The boundary conditions induce highly nonlocal effects
in the bulk. One effect of the boundary conditions is \cite{Zinn00}
that one has phase separation. It is not obvious to see such a phase
separation if one looks at the RSOS interface. 

The various aspects of ice model with special boundary
conditions are related to the ``The many faces of the alternating-sign
matrices'' \cite{Propp01} to which, we think that the RPM has added a
useful new one. For example, the RPM model allows for easy Monte-Carlo
simulations in order to obtain correlation functions, which are much
harder to get for the other ``faces'' of the alternating sign matrices.    

We now sum up the main results we have obtained about the physical
properties of the RPM for $u=1$. In  Section \ref{se:heights} we show
that the average height and width of the interface increase
logarithmically with the system size indicating that the interface is 
only marginally rough, $3/4$ of the interface being covered by terraces.

Based on a conjecture for the expression of the probability to have $k$
clusters ($k+1$ contacts of the interface with the horizontal line)
\cite{Gier02} for a system of size $L$ and introducing a fugacity
(enhancement parameter) $\zeta$ related to the number of contacts, we have
derived in Section \ref{se:clusters} the interface tension as a
function of $\zeta$. An important result of this paper is that there
is a special surface transition at $\zeta=1$,
separating a phase ($\zeta<1$) with a finite number of clusters (density
zero) from a phase ($\zeta>1$) with a finite density of clusters. This
implies that no enhancement ($\zeta>1$) is needed to obtain the special surface
transition \cite{Bind83}. This fact which is certainly related to the unusual role
of the boundaries in the FPL model, is interesting for two reasons. Firstly we
are not aware of any model in which one obtains a special transition without
enhancement. Secondly, the RPM is related to the dense O(1) model
($q=\exp(\i\pi/3)$ in eq. (\ref{eq:TLdef})). The common lore
\cite{Burk89} is that the ordinary O($n$) model can have a special
transition with a finite enhancement for $-2 \leq n<1$ only. We not only 
have the surface transition but we get it without enhancement. Various
scaling laws have been checked and the crossover exponent $\phi=2/3$ was obtained.  
This exponent also gives the large $L$ behavior of the susceptibility
related to the two-point contacts correlation function. This correlation
can be obtained for example by using Monte-Carlo simulations in the
RPM. It would be much more difficult to obtain it directly in the
FPLM. 

In Section \ref{se:ava} we have studied the SOC properties of the RPM
showing that the avalanches of tiles produced by desorption have a PDF
which, for large systems sizes, has a finite average (as opposed to
sand pile models) but a divergent dispersion.

In Section \ref{se:time} we have studied the RPM for $u$ different from $1$. No
connections with FPLM or with integrable systems are known in this case. For
small and very large values of $u$ the system is massive. Finite size
scaling studies suggest that around $u=1$ the dynamic scaling exponent
$z$ has a value closed to $1$ (for $u=1$ one has
$z=1$) but that conformal invariance is lost. Such a scenario was not 
seen in other models. 

The list of open questions which should start with the possible
physical applications of the model is too long to be written. Progress in
combinatorics (to prove the conjectures) and efforts to obtain analytic
expressions for the correlation functions are obviously required. We
assume that the first results which will push forwards the
understanding of the model will come from Monte-Carlo simulations.
\section*{Acknowledgements} 
This research is supported by the 
Australian Research Council, and by the Foundation `Fundamenteel
Onderzoek der Materie' (FOM). VR is an ARC IREX Fellow. We thank
H. Hinrichsen and A. Owczarek for discussions.   

\appendix

\section{Physical interpretation of the asymptotics of $A^{\rm
V}_{2n+1}$}

We now show that the surface contribution to the asymptotics of
$A^{\rm V}_{2n+1}$, given in (\ref{eq:AVasymp}), has the
interpretation of a probability. To see that we compare
(\ref{eq:AVasymp}) to the asymptotics \cite{BogoKZ02} of the number
$A_{L+1}$ \cite{Zeilb96b,Kupe96} of six-vertex configurations on the
$(L+1)\times (L+1)$ square with domain wall boundary conditions. This
expression does not contain a surface term,
\be
\log A_{L+1} = s_0 (L+1)^2 -\frac{5}{72} \log L^2 +O(1).
\label{eq:Aasymp}
\ee
We stress that a surface term does appear when one considers weighted
configurations of the six-vertex model \cite{Zinn00}.  

\begin{figure}[ht]
\centerline{
\begin{picture}(240,80)
\put(0,0){\epsfxsize=80pt\epsfbox{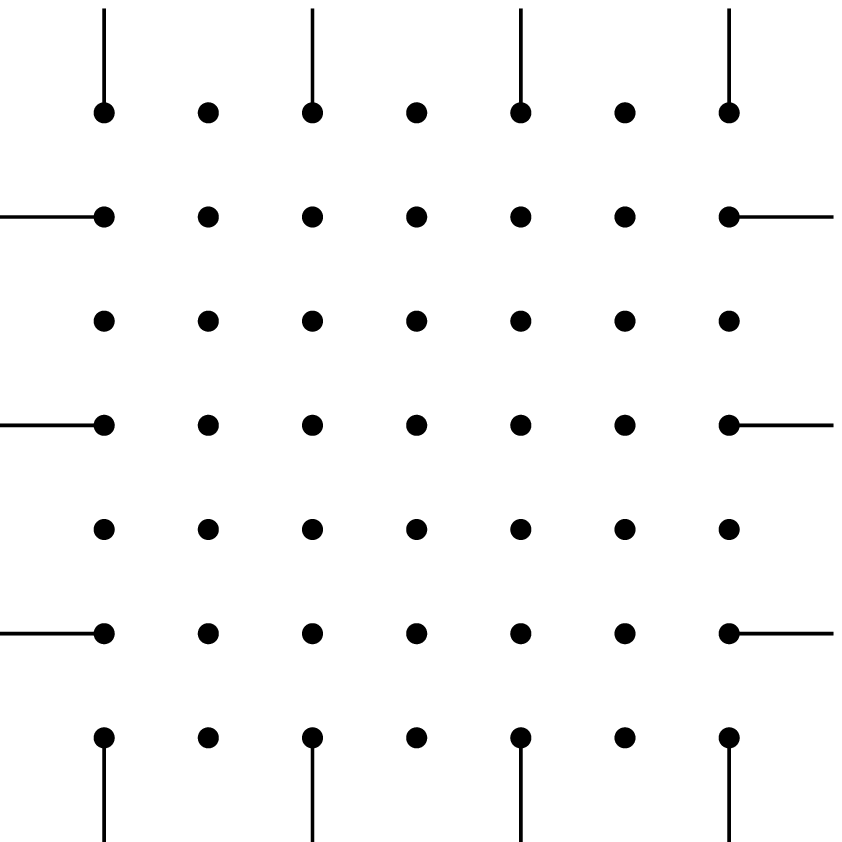}}
\put(120,40){$\rightarrow$}
\put(160,0){\epsfxsize=80pt\epsfbox{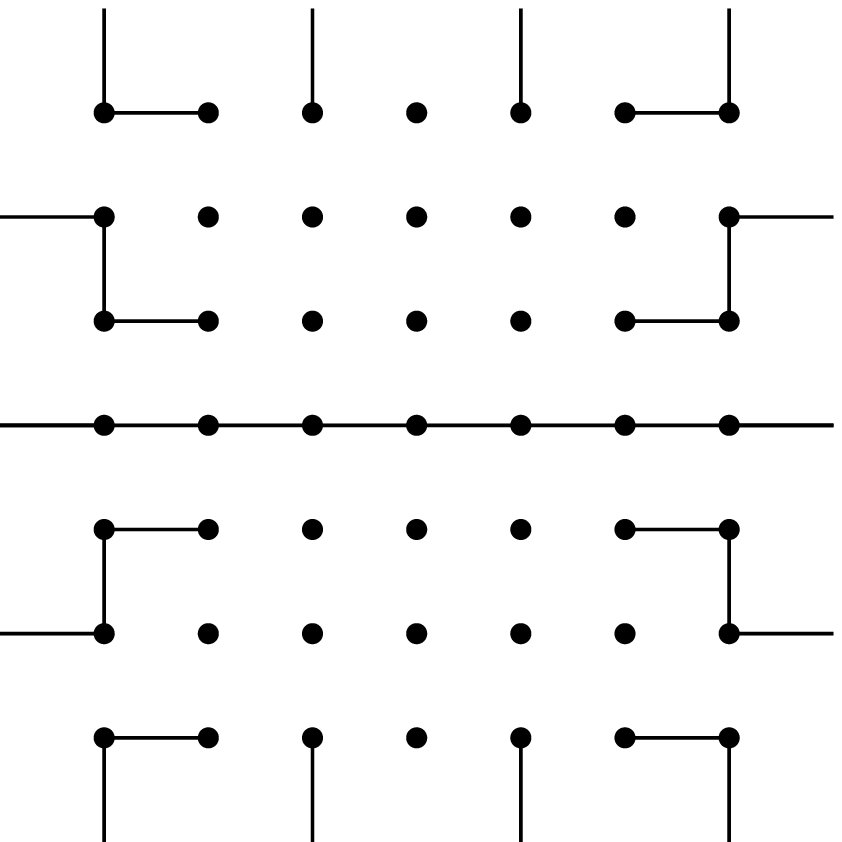}}
\end{picture}}
\caption{Fixed edges for horizontally symmetric FPL diagrams.} 
\label{fig:fixed}
\end{figure}

To understand the meaning of the surface contribution in
(\ref{eq:AVasymp}) we use the FPLM. Requiring horizontal symmetry for
FPL diagrams on a square implies that a loop segment runs along the
horizontal symmetry axis and also that the left and right boundary
layers are fixed, see Fig. \ref{fig:fixed}. The number $A^{\rm
V}_{L+1}$ thus counts all FPL diagrams on the lower (or upper)
rectangle of the right hand side picture of Fig. \ref{fig:fixed}.

The probability that the horizontal symmetry axis of an arbitrary FPL
diagram on the square is covered by a loop segment can now be 
easily calculated. Using (\ref{eq:AVasymp}) and (\ref{eq:Aasymp}) we find,
\be
\frac{\left(A^{\rm V}_{2n+1}\right)^2} {A_{2n+1}} \sim
\left( \frac12\e^{-s_0} \right)^{2n} \qquad (n
\rightarrow\infty),
\label{eq:Vprob}
\ee
where $s_0$ is the entropy per site given in (\ref{eq:AVasymp}). Note that
(\ref{eq:Vprob}) is a product of probabilities in which the value $1/2$
is screened by a factor $\e^{-s_0}$ due to bulk effects. The
probability in (\ref{eq:Vprob}) can be interpreted as a string
expectation value (see e.g. \cite{KoreBI93}).

The surface contribution in (\ref{eq:AVasymp}) therefore has the
interpretation of a probability. It arises due to the boundary with
the alternating arrows in Fig. \ref{fig:dwbcex}, and not from the
domain wall boundaries on the sides and bottom. The fact that the
leading behaviour in (\ref{eq:Vprob}) is purely exponential and does
not contain an algebraic factor is very surprising for the following reason. 
The probability that the first edge of the horizontal symmetry axis is
drawn is known as the refined alternating sign matrix conjecture
\cite{MillsRR82} which meanwhile is proved \cite{Strog02,Zeilb96}. We
denote this probability by $p_1$ and it is given by,
\be
p_1= 
\binom{3n-1}{n-1} \frac{\Gamma(3n+2)\Gamma(4n+2)}{\Gamma(n+2)\Gamma(6n+2)}
\approx
\sqrt{\frac{2}{3\pi n}},
\ee
which is algebraic. In the probability (\ref{eq:Vprob}) that the
entire horizontal symmetry axis contains a loop segment, the algebraic
dependence has disappeared and is transformed into a screening factor
of $\e^{-s_0}$ per edge. A further amusing point is that for $n=1$,
(\ref{eq:Vprob}) gives an esitmate of $0.2798...$ for the entropy 
per site which is already very close to its exact value $s_0 =
0.2616...$.


\begin{thebibliography}{99}
%
\bibitem{BaraS95} Some recent reviews are A-L. Barabasi and H.E. Stanley, 
{\it Fractal concepts in surface growth}, Cambridge University Press, 1995;
E.H. Hinrichsen, Adv. Phys. {\bf 49}, 815 (2000).

\bibitem{GierNPR02} J. de Gier, B. Nienhuis, P.A. Pearce and
V. Rittenberg, {\it Phys. Rev. E} {\bf 67}, 016101 (2002).

\bibitem{BakTW87} P. Bak, C. Tang and K. Wiesenfeld, Phys. Rev. Lett. 
{\bf 59}, 381 (1987); {\it J. Phys. A} {\bf 38}, 364 (1988).

\bibitem{Jensen98} Some recent reviews are H. J. Jensen, {\it Self 
Organised Criticality}, Cambridge University Press, 1998;
D. Dhar, Physica A 264, 1 (1999).

\bibitem{BenHur96} A. Ben-Hur and O. Biham, Phys. Rev. E {\bf 53},
R1317 (1996); A. Vespignani and S. Zapperi, Phys. Rev. E {\bf 57},
6345 (1998).

\bibitem{BatchGN01} M. T. Batchelor, J. de Gier and B. Nienhuis, J. 
Phys. A {\bf 34}, L265 (2001).

\bibitem{PearceRGN02} P. A. Pearce, V. Rittenberg, J. de Gier and
B. Nienhuis, {\it J. Phys. A} {\bf 35}, L661 (2002).

\bibitem{AlcarazBBBQ} F. C. Alcaraz, M. N. Barber, M. T. Batchelor,
R. J. Baxter and G. R. W. Quispel, J. Phys. A {\bf 20}, 6397 (1987).

\bibitem{KogN02} I.I. Kogan and A. Nichols, arXiv:hep-th/0203207 and 
references therein; Flohr M 2001 Bits and pieces in logarithmic
conformal field theory {\it Preprint} arXiv:hep-th/0111228; Gaberdiel
M R 2001 An algebraic approach to logarithmic conformal field theory
{\it Preprint} arXiv:hep-th/0111260; Kawai S 2002 Logarithmic conformal
field theory with boundary {\it Preprint} arXiv:hep-th/0204169.

\bibitem{GurL99} V.Gurarie and A.W.W. Ludwig, arXiv:cond-mat/9911392;
I.I. Kogam and A.M. Tsvelik, Mod. Phys. Lett. A {\bf 15}, 931 (2000).

\bibitem{SchouF02} K. Schoutens and P. Fendley, arXiv:hep-th/0210161.

\bibitem{KogP01} I.I. Kogan and D. Polyakov, Int. J. Mod. Phys. A {\bf
16}, 2559 (2001).

\bibitem{Robbins00}
D.P. Robbins, {\it Symmetry classes of alternating sign matrices}, (2000);
arXiv:math.CO/0008045.

\bibitem{Kupe00} G. Kuperberg, math.CO/0008184 (2000);

\bibitem{Bress99} Bressoud D M 1999 {\it Proofs and Confirmations: The
story of the Alternating Sign Matrix Conjecture} (Cambridge: Cambrige
University Press) 

\bibitem{Lieb67} E. Lieb, Phys. Rev. Lett. {\bf 18}, 692 (1967).

\bibitem{ElkiesKLP92} N. Elkies, G. Kuperberg, M. Larsen, and
J. Propp, J. Algebraic Combin., 1 (1992), pp. 111-132 and 219-234.
%
\bibitem{BatchBNY96} M. T. Batchelor, H. W. J. Bl\"ote, B. Nienhuis and
C. M. Yung, {\it J. Phys. A} {\bf 29}, L399 (1996); 

\bibitem{Wiel00} B. Wieland, {\it Electron. J. Combin.} {\bf 7},
research paper 37 (2000).

\bibitem{RazuS01} A. V. Razumov and Yu. G. Stroganov, math.CO/0104216
(2001), math.CO/0108103 (2001); 

\bibitem{Zinn00} P. Zinn-Justin, {\it Phys.Rev. E} {\bf 62} (2000),
341; V. Korepin and P. Zinn-Justin, {\it J.Phys. A} {\bf 33} (2000),
7053; P. Zinn-Justin, arXiv:cond-mat/0205192.

\bibitem{Propp01} J. Propp, {\it Discr. Math. and
Theor. Comp. Sci. Proc. AA}, 43 (2001).

\bibitem{BogoPZ02} N.M. Bogoliubov, A.G. Pronko and M.B. Zvonarev,
{\it J. Phys. A} {\bf 35}, 5525 (2002).

\bibitem{Strog02} Yu.G. Stroganov, arXiv:math-ph/0204042.

\bibitem{Bind83} K. Binder, in {\it Phase Transitions and Critical
Phenomena} {\bf 8}, eds. C. Domb and J.L. Lebowitz, Academic Press
Inc. (London), 2 (1983).

\bibitem{Balian73} R. Balian and G. Toulouse, {\it Ann. Phys.} {\bf
83}, 28 (1973).

\bibitem{Burk89} T.W. Burkhardt, E. Eisenriegler and I. Guim, {\it
Nucl.Phys. B} {\bf 316} (1989), 559.

\bibitem{Queiroz95} S.L.A. Queiroz, {\it J. Phys. A} {\bf 27}, L363
(1995).

\bibitem{Ruelle02} P. Ruelle, private communication.

\bibitem{Ruelle01} S. Mathieu and P. Ruelle, {\it Phys.Rev. E} {\bf
64} (2001), 066130; P. Ruelle, arXiv:hep-th/0203105 and references
therein.

\bibitem{IvashP98} E.V. Ivashkevich, V.B. Priezzhev, {\it Physica A}
{\bf 254}, 97 (1998).

\bibitem{Dhar99} D. Dhar, {\it Physica A} {\bf 263}, 4 (1999).

\bibitem{Rouse53} P.E. Rouse, {\it J.Chem.Phys.} {\bf 48} (1953) 57.

\bibitem{Priv80} V. Privman and N.M. Svrakic, {\it Directed models of polymers, Interfaces,
and clusters: scaling and finite size properties}, Lecture Notes in
Physics {\bf 38}, Springer Verlag, Berlin Heidelberg 1989.

\bibitem{KodD98} H.N. Koduvely and D.Dhar, {\it J. Stat. Phys.} {\bf
90} (1998), 57.

\bibitem{Pasq87} V. Pasquier, {\it Nucl. Phys. B} {\bf 285}, 162
(1987); {\it J. Phys. A} {\bf 20}, L1229 (1987); {\it J. Phys. A} {\bf
20}, 5707 (1987).

\bibitem{OwczaB87} A. Owczarek and R.J. Baxter, {\it J. Stat. Phys.}
{\bf 49}, 1093 (1987).

\bibitem{Mart91} P.P. Martin, {\it Potts models and related
problems in statistical mechanics}, World Scientific
Singapore (1991).

\bibitem{PasqS90} V. Pasquier and H. Saleur, Nucl. Phys. {\bf B 330}, 
523 (1990).

\bibitem{Kore82} V. E. Korepin, Comm.\ Math.\ Phys. {\bf 86}, 391
(1982);

\bibitem{MillsRR82} W.H. Mills, D.P. Robbins and H. Rumsey,
Invent. Math. {\bf 66}, 73 (1982); W.H. Mills, D.P. Robbins and
H. Rumsey, J. Combin. Theory Ser. A {\bf 34}, 340 (1983).  

\bibitem{Zeilb96} D. Zeilberger, {\it New York J. Math.} {\bf 2}, 59
(1996).

\bibitem{Strog} We thank Yu. G. Stroganov for this conjecture.

\bibitem{HinrS02} H. Hinrichsen and L. Sittler, private communication.

\bibitem{Fam85} F. Family and E. Vicsek, {\it J. Phys. A} {\bf 18},
L75 (1985); {\it Dynamics of Fractal Surfaces}, World Scientific
Singapore, 1991. 

\bibitem{AlonEHM98} U. Alon, M. R. Evans, H. Hinrichsen and D. Mukamel,
Phys. Rev. E {\bf 57}, 4997 (1998).

\bibitem{Hinri01} H. Hinrichsen, private communication.

\bibitem{Brak98} R. Brak, J.W. Essam and A.L. Owczarek, {\it
J.Stat.Phys.} {\bf 93} (1998), 155;  {\it J.Stat.Phys.} {\bf 102}
(2001), 997, and references therein. 

\bibitem{Gier02} J. de Gier, arXiv:math.CO/0211285.

\bibitem{Bateman} A. Erd\'elyi, {\it Higher transcendental functions}
{\bf 1}, McGraw-Hill, New York 1953.

\bibitem{Gennes79} P.G. de Gennes, {\it Scaling concepts in polymer physics}, Cornell
Univ. Press, Ithaca, NY 1979.
%
\bibitem{PrellO94} T. Prellberg and A.L. Owczarek, {\it J.Phys. A}
{\bf 27} (1994) 1811.

\bibitem{Zande98} C. Vanderzande, {\it Lattice Models of polymers},
Cambridge University Press, Cambridge 1998.

\bibitem{Pol70} V. Polyakov, {\it Zh. Eksp. Theor. Fiz.} {\bf 59}
(1970), 542.

\bibitem{Cloiz90} J. des Cloizeaux and Gerard Jannink, {\it Polymers
in Solution}, Clarendon Press Oxford, 1990.

\bibitem{Alex} A.L. Owczarek (private communication).

\bibitem{PauleS95} P. Paule and M. Schorn, {\it J. Symb. Comp.} {\bf 20},
673 (1995).

\bibitem{Batch93} M.T. Batchelor, {\it J.Phys. A} {\bf 26} (1993), 3733.

\bibitem{Dupl87} B. Duplantier and H. Saleur, {\it Phys. Rev. Lett.}
{\bf 59} (1987), 539.

\bibitem{ReadS01} N. Read and H. Saleur, {\it Nucl. Phys. B} {\bf
613}, 409 (2001), and references therein.

\bibitem{Tebaldi99} C. Tebaldi, M. De Menech and A. L. Stella,
{\it Phys. Rev. Lett.} {\bf 83}, 3952 (1999). 

\bibitem{VBS} J.M. van den Broeck and L.W. Schwartz, Siam
J. Math. Anal. {\bf 10}, 639 (1979).

\bibitem{BatchGN02} M. T. Batchelor, J. de Gier and B. Nienhuis, {\it
Int. J. Mod. Phys. B} {\bf 16}, 1883 (2002). 

\bibitem{BogoKZ02} N.M. Bogoliubov, A. V. Kitaev and M. B. Zvonarev,
{\it Phys. Rev. E} {\bf 65}, 026126 (2002).

\bibitem{Zeilb96b} D. Zeilberger, {\it Electr. J. Combin.} {\bf 3},
R13 (1996). 

\bibitem{Kupe96} G. Kuperberg, {\it Invent. Math. Res. Notes} {\bf
1996}, 139 (1996).

\bibitem{KoreBI93} V.E. Korepin, N.M. Bogoliubov and A.G. Izergin,
{\it Quantum inverse scattering method and correlation functions},
Cambridge University Press, Cambridge (1993).
%
\end{thebibliography}
\end{document}